\pdfoutput=1
\documentclass[11pt]{article}

\usepackage[margin=1in]{geometry}
\usepackage{microtype}

\usepackage{graphicx}

\usepackage{amsmath}
\usepackage{amssymb}

\usepackage{subcaption}
\usepackage{caption}

\usepackage{xcolor}
\usepackage{tikz}
\usetikzlibrary{shapes}

\usepackage[pagewise,displaymath,mathlines]{lineno}

\usepackage[numbers,sort&compress]{natbib}
\usepackage[colorlinks=true,linkcolor=blue,citecolor=blue,urlcolor=blue]{hyperref}

\newcommand{\redx}[1]{{\color{red}#1}}
\newcommand{\bluex}[1]{{\color{blue}#1}}
\newcommand{\greenx}[1]{{\color{green}#1}}
\newcommand{\tikzcircle}[2][red,fill=red]{%
  \tikz[baseline=-0.5ex]\draw[#1,radius=#2] (0,0) circle ;%
}

\title{Scalar Dispersion from Wall-Mounted Cylinders at Large Reynolds Number: Plume Transitions and Regime Classification}

\author{
\begin{tabular}{c}
Kofi Agyemang Amankwah$^{1}$,
Juan Carlos Cuevas Bautista$^{1}$,
Theresa Oehmke$^{1}$,\\
Christopher M. White$^{1}$\thanks{Corresponding author: \href{mailto:chris.white@unh.edu}{chris.white@unh.edu}},
Lukasz Zielinski$^{2}$,
Gocha Chochua$^{2}$,
Andrew Speck$^{2}$\\[0.6em]
{\small $^{1}$Department of Mechanical Engineering, University of New Hampshire, Durham, NH 03824, USA}\\
{\small $^{2}$Schlumberger-Doll Research, One Hampshire St, Cambridge, MA 02139, USA}
\end{tabular}
}

\date{} 

\begin{document}
\maketitle


\begin{abstract}
This study presents a comprehensive experimental investigation of scalar dispersion from the free end of wall-mounted cylindrical obstacles immersed in a large-Reynolds-number turbulent boundary layer. A key focus is the characterization of transition behavior between distinct dispersion regimes: elevated plumes (EP), ground-level plumes (GLP), and ground-level sources (GLS). Experiments systematically vary the primary and secondary aspect ratios (\( AR_1 \), \( AR_2 \)) and the velocity ratio (\( r \)) to explore their effects on the evolution of scalar plumes. Plume classification is governed by the non-dimensional parameter \(\tilde{h}_s / \delta_{\mathrm{cz}}\), which quantifies the progressive interaction between the plume and the ground. Here, \(\tilde{h}_s\) denotes the effective source height and \(\delta_{\mathrm{cz}}\) the vertical plume half-width. Detailed concentration measurements demonstrate that the EP--GLP--GLS transitions substantially modify both vertical and lateral dispersion characteristics. The measurements reveal systematic departures from classical dispersion-coefficient scaling. To assess the capability of existing models under these conditions, the experimentally determined dispersion coefficients are used to evaluate the Gaussian Dispersion Model (GDM) and a Wall Similarity Model (WSM). The GDM captures general trends but deviates in specific regimes, whereas the WSM offers improved representation under GLS conditions. The resulting dataset, grounded in systematic laboratory measurements, establishes a critical benchmark for validating numerical simulations and informing the development of next-generation predictive models. Finally, leveraging these results, a concise data-informed predictive framework is introduced that captures the EP--GLP--GLS transitions and provides first-order estimates of ground-level concentration across geometric and momentum-ratio parameter space.
\end{abstract}



\section{Introduction}

The ability to accurately predict scalar dispersion in turbulent boundary layer flows influenced by three-dimensional wall-flow effects is of critical importance across a wide range of natural processes and engineering and industrial applications. These include scalar transport around submerged obstacles in tidal flows, pollutant release from industrial facilities into the atmosphere, and the release of radionuclides, toxic chemicals, or airborne pathogens in urban environments. A fundamental challenge in accurately predicting scalar dispersion from a point source within the atmospheric boundary layer lies in the complex flow interactions induced by ground-based obstacles, typically situated at or downstream of the source and fully immersed within the boundary layer \citep{Meroney1990}. These obstacles generate inherently three-dimensional, unsteady flow fields characterized by concentrated regions of vorticity, complex vortex–vortex interactions, and high levels of fluctuating vorticity. Such coherent and incoherent turbulent flow structures can significantly influence, and in some cases intermittently dominate, scalar transport, leading to highly anisotropic and non-Gaussian dispersion behavior.

Owing to these complexities and their critical implications for environmental quality and public health, scalar dispersion downwind of wall-mounted obstacles immersed in boundary layer flow has long been, and remains, an active area of research \citep{Allwine1980, Hunt1984, Thompson1993, Castelli2010, Miao2014, Blocken2018, Mo2018}. These studies aim to advance our understanding of the underlying flow physics that govern scalar transport in such configurations, with the goal of developing more accurate predictive models for use in urban environments and complex industrial settings. Dispersion models used for regulatory compliance are based on Gaussian dispersion models (GDM) \citep{Schulman2000, Cimorelli2005}. The modeling and analysis then focus on predicting dispersion coefficients as a function of downwind distance, under a range of flow conditions, including the influence of wakes generated by ground-based structures. In contrast, simulation-based approaches, such as RANS, LES, and DNS, are more computationally intensive but can offer greater accuracy and insight \citep{Sada2002, Tseng2006, Shi2008, Rossi2009, Xie2009, Castelli2010, Blocken2018}.

In this study, scalar dispersion is investigated for releases from the free end of finite-height, wall-mounted cylinders immersed in a developing turbulent boundary layer at high Reynolds number. A near-neutrally buoyant mixture of 5\% methane and 95\% nitrogen by volume is injected at mean velocity \(u_s\) through a circular source of diameter \(d_s\) centered on the cylinder free end. The cylinder height \(h_s\) is held fixed, yielding \(1.45 \le \delta/h_s \le 1.81\), where \(\delta\) is the local boundary-layer thickness. Four release geometries are examined, characterized by primary aspect ratio \(AR_1 \equiv h_s/d_o\) and secondary aspect ratio \(AR_2 \equiv d_o/d_s\): \(AR_1 = 11.47,\, 4.72,\, 1.3,\, 0.7\) and \(AR_2 = 1.22,\, 2.96,\, 10.75,\, 19.95\). To the authors’ knowledge, these experiments represent the highest boundary-layer Reynolds number (\(\delta^+ \simeq 6{,}210\)) and bluff-body Reynolds number (\(Re_{d_o} \simeq 1.82\times10^5\)) reported for wind-tunnel studies of scalar dispersion involving wall-mounted obstacles.

The primary objective is to quantify how three-dimensional wall-flow interactions govern EP–GLP–GLS plume transitions, which strongly influence ground-level exposure. In this context, the predictive performance of the Gaussian Dispersion Model (GDM) is assessed under transition conditions and compared against a Wall Similarity Model (WSM). The resulting high-Reynolds-number dataset provides validation material for numerical simulations and analytical dispersion models relevant to complex near-wall release scenarios. Finally, leveraging these results, a concise data-informed predictive model is introduced that discriminates EP–GLP–GLS transitions and provides first-order estimates of ground-level concentration across geometric and momentum-ratio parameter space.

\section{Scalar dispersion models}

\subsection{Gaussian dispersion model} \label{GDM}

The Gaussian dispersion model underpins most open-air pollutant models employed by regulatory agencies, including the U.S. EPA’s AERMOD \citep{Cimorelli2005}. While the primary advantage of GDM lies in its simplicity and tractability, this comes at the cost of generalized assumptions that can be violated in complex wall-bounded flows \citep{Veigele1978}. Despite these limitations, the GDM often performs well in practice when empirically calibrated parameters are employed \citep{hanna2009, dou2022, sofia2020}.

The standard GDM formulation of \citet{Pasquill1974} to compute steady concentration \(C(x,y,z)\) is recast into a form that is more convenient for describing and interpreting the EP--GLP--GLS transition behavior:
\begin{align}
\begin{split}
C(x,y,z) &= \frac{Q \ln(2)}{\pi U \, \delta_{\mathrm{cy}} \delta_{\mathrm{cz}}}
\exp\!\Bigg[-\ln(2)\frac{(y - y_s)^2}{\delta_{\mathrm{cy}}^2}\Bigg] \\
&\quad \times 
\Bigg\{
\exp\!\Bigg[-\ln(2)\frac{(z - \tilde{h}_s)^2}{\delta_{\mathrm{cz}}^2}\Bigg]
\Big[ 1 + \exp\!\Big(-\ln(2)\frac{4z\tilde{h}_s}{\delta_{\mathrm{cz}}^2}\Big) \Big]
\Bigg\},
\end{split}
\label{eq:GDM2}
\end{align}
for a source located at \((0, y_s, h_s)\) with emission rate \(Q\) in the presence of a wall at \(z = 0\). Here, \(\delta_{\mathrm{cy}}\) and \(\delta_{\mathrm{cz}}\) denote the plume half-widths at half-maximum (HWHM), \(U\) is the steady downwind velocity, and \(\tilde{h}_s = h_s + \Delta h_s\) is the effective source height accounting for plume rise or fall due to buoyancy or source effects. The image-source term \((\exp[-\ln(2)\,4z\tilde{h}_s / \delta_{\mathrm{cz}}^2])\) enforces the boundary condition \(\partial C / \partial z |_{z=0} = 0\).

\begin{figure}
\centering
  \includegraphics[scale=0.25]{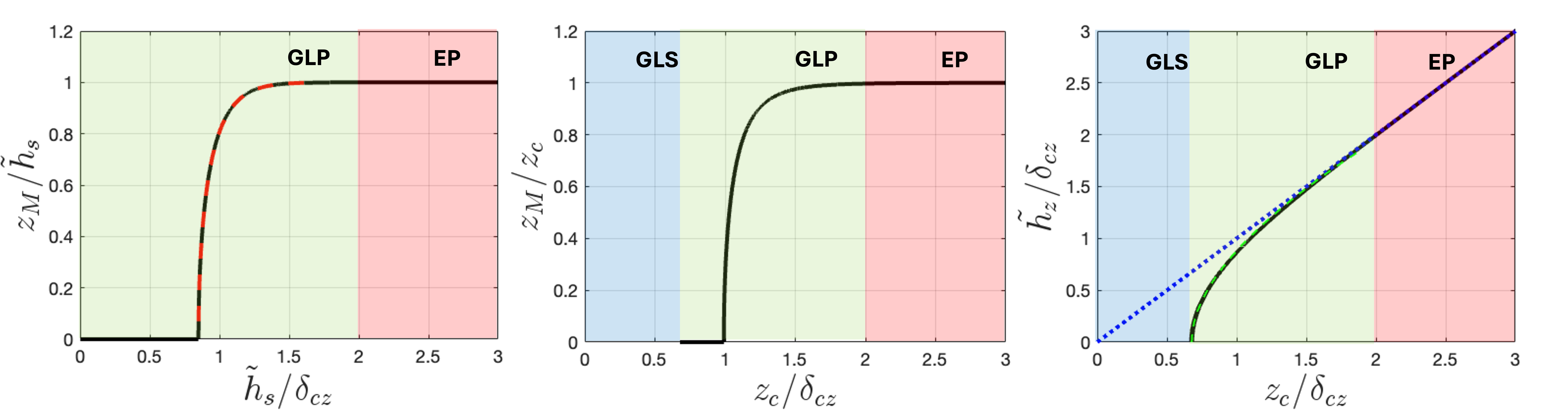}
\caption{\small{(left) The normalized vertical location of peak concentration $z_M/\tilde{h}_s$ versus $\tilde{h}_s/\delta_{\mathrm{cz}}$ (black curve). The red dashed line is a derived analytical solution \(\tilde{h}_s/\delta_{\mathrm{cz}} = \sqrt{ \tanh^{-1}(\hat{z}_M)/( 2 \ln(2) \, \hat{z}_M)}\) valid for $0 < z_M/\tilde{h}_s < 1$, where $\hat{z}_M = z_M/\tilde{h}_s$. (middle) $z_M/z_c$ versus $z_c/\delta_{\mathrm{cz}}$. (right) $\tilde{h}_s/\delta_{\mathrm{cz}}$ versus $z_c/\delta_{\mathrm{cz}}$. The green dotted line is the fit $\tilde{h}_s/\delta_{\mathrm{cz}} = 1.68 (z_c/\delta_{\mathrm{cz}} - \chi_\text{GLS})^{0.66}+ 0.05(z_c/\delta_{\mathrm{cz}} - \chi_\text{GLS})$, valid for $0 < z_c/\tilde{h}_s < 1.55$.} The blue dotted line is $\tilde{h}_s/\delta_{\mathrm{cz}}$ = $z_c/\delta_{\mathrm{cz}}$ shown for reference.}
\label{fig:GLP}
\end{figure}

The parameter that distinguishes EP, GLP, and GLS is \(\tilde{h}_s/\delta_{\mathrm{cz}}\). Plume classification metrics that are height-dependent but independent of other release-geometry parameters—\(z_M/\tilde{h}_s\), \(z_c/\delta_{\mathrm{cz}}\), and their mapping—are summarized in figure~\ref{fig:GLP}, where \(z_M\) denotes the peak-concentration height and \(z_c\) the plume centroid. EPs interact weakly with the wall and are categorized by \(\tilde{h}_s / \delta_{\mathrm{cz}} \geq 2\). GLPs are categorized by \(0 < \tilde{h}_s / \delta_{\mathrm{cz}} < 2\), with the image-source contribution increasing as \(\tilde{h}_s / \delta_{\mathrm{cz}}\) decreases. At \(\tilde{h}_s / \delta_{\mathrm{cz}} = 0.85\), the peak concentration shifts to the wall, signifying a transition from dominance by the real source to dominance by the combined real–image source. The GLS regime is categorized by \(\tilde{h}_s / \delta_{\mathrm{cz}} = 0\) and, as derived from (\ref{eq:GDM2}) or more generally from (\ref{eq:Cprime2}), \((z_c / \delta_{\mathrm{cz}})_\text{GLS} = B / (\ln 2)^{1/s} \approx 0.68\), where \(B(s) = \Gamma(2/s) / \Gamma(1/s)\), \(\Gamma\) denotes the gamma function, and \(s = 2\). Herein, \(\chi_{GLS} \equiv (z_c / \delta_{\mathrm{cz}})_\text{GLS}\).

Direct measurement of \(\tilde{h}_s\) within the GLP regime is not feasible (here \(\tilde{h}_s \neq z_c\); for example, in a GLS configuration \(\tilde{h}_s = 0\) while \(z_c > 0\)). Instead, the plume centroid height \(z_c\) provides a measurable surrogate. The right-hand panel of figure~\ref{fig:GLP} shows the relationship between \(z_c/\delta_{\mathrm{cz}}\) and \(\tilde{h}_s/\delta_{\mathrm{cz}}\). The two quantities coincide closely until \(\tilde{h}_s/\delta_{\mathrm{cz}} \lesssim 1.5\). The green dotted line denotes the empirical fit used to map \(z_c/\delta_{\mathrm{cz}}\) to \(\tilde{h}_s/\delta_{\mathrm{cz}}\).

\subsection{Wall similarity models}
\label{sec:WSM}

Serving as a comparator to the GDM, the WSM is examined in the context of plumes transitioning from a GLP to a GLS. WSM incorporates height-dependent mean velocity and diffusivity, e.g. \(U(z)=u_1 z^m\), \(D_z(z)=K_1 z^n\) \citep{vanulden1978,Gavze2018}. For a GLS, the WSM crosswind \(y\)-integrated profile \(C^\prime=\int C\,dy\) has the normalized form 
\begin{equation}
\frac{C^\prime(x,z)}{C^\prime_M} = \exp \!\left[ -\ln 2 \left(\frac{z}{\delta_{\mathrm{cz}}}\right)^s \right],
\label{eq:Cprime2}
\end{equation}
with shape factor \(s=2+m-n\); \(s=2\) recovers a Gaussian distribution. Field and laboratory studies generally find \(1.5\lesssim s \lesssim 1.7\): \citet{Robins1978} reported 
\(s\) = 1.7 in urban flows; \citet{Fackrell982} reported \(s\) = 1.5 in zero pressure gradient (ZPG) rough-wall boundary layer; \citet{Lim2023} reported \(s\) = 1.5 in ZPG smooth-wall boundary layer. The non-Gaussian behavior implies that for \(s < 2\),  
\(\chi_{\mathrm{GLS,WSM}} > \chi_{\mathrm{GLS,GDM}}\); for example, when \(s = 1.5\),  
\(\chi_{\mathrm{GLS,WSM}} \approx 0.84\).  Thus, defining the GLP–GLS transition solely by \(\chi_{\mathrm{GLS}}\) is not without uncertainty.

\section{Experiments}\label{sec:rules_submission}
\subsection{Facility and scalar release geometry}

\begin{figure}
\centering
\includegraphics[scale=0.4]{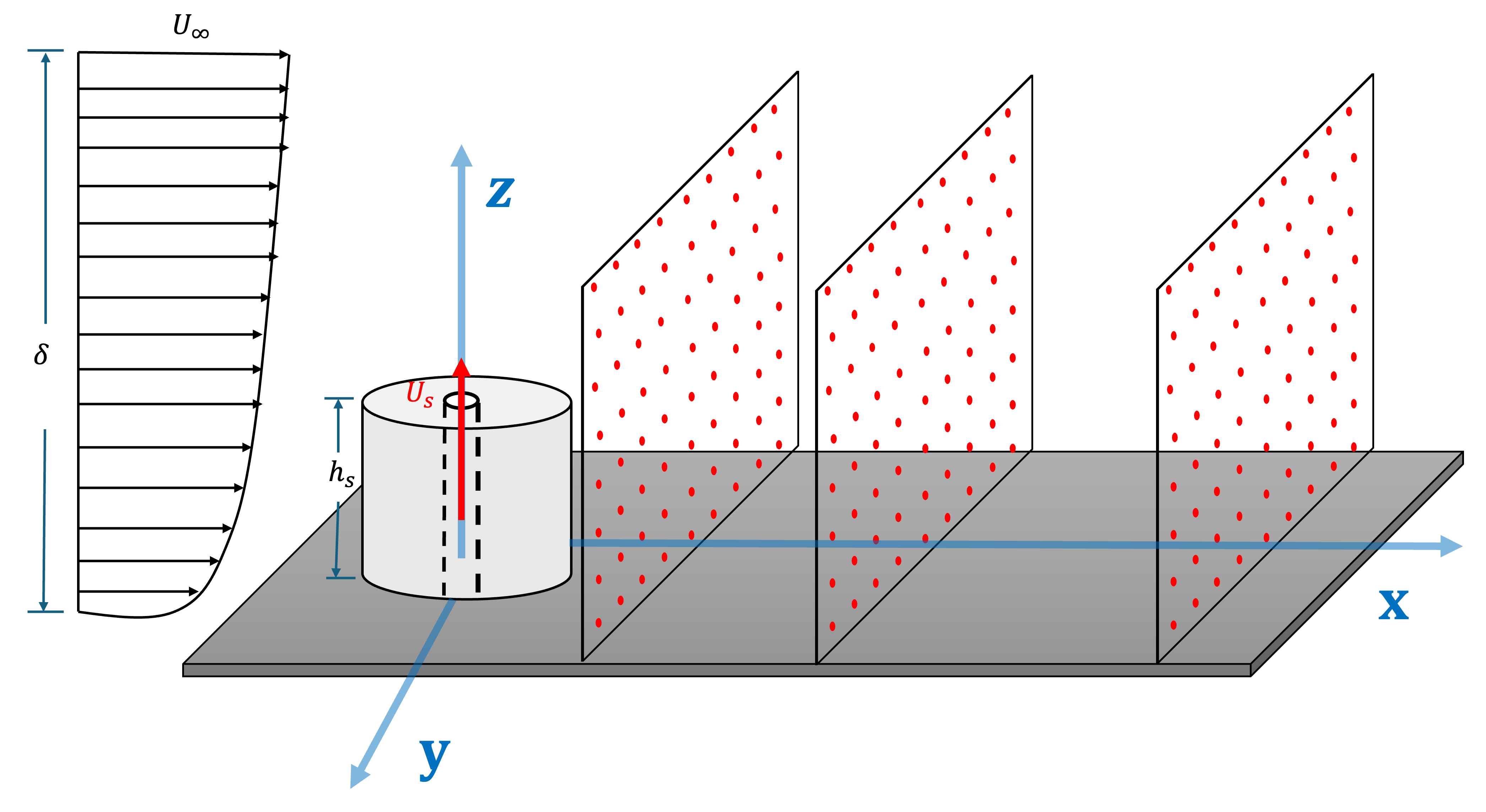}
\caption{\small Schematic of the scalar release setup (GEOM2–GEOM4 shown). The inner dashed cylinder is the source (GEOM1). Measurement planes \(P1,P2,P3\) capture plume development.}
\label{fig:setup}
\end{figure}

Experiments are performed in the UNH Flow Physics Facility (FPF), a ZPG wind tunnel with \(2.8\times6\,\mathrm{m}^2\) cross-section and 72\,m fetch; see \citet{vincenti2013}. The setup is shown in figure~\ref{fig:setup}. Four wall-mounted cylinders share a common internal source (GEOM1) and differ by outer diameter. Geometry details and aspect ratios \(AR_1=h_s/d_o\), \(AR_2=d_o/d_s\) are in Table~\ref{table:LG}. GEOM1 consists of a smooth-walled polyvinyl chloride (PVC) cylinder pipe with an inner diameter of \( d_s = 20.375 \pm 0.056\,\mathrm{mm} \) and an outer diameter of \( d_o = 24.85 \pm 0.055\,\mathrm{mm} \). The ends are precisely machined to be flat and sharp-edged. This pipe is vertically inserted into the FPF test section at \(x_s = 27.54\) m through an aperture in the test section floor. The three additional release geometries are constructed using larger-diameter PVC pipes, also machined with flat surfaces and sharp edges. A smooth free end surface is machined to tolerance from 19.05\,mm-thick high-density polyethylene (HDPE) and press-fitted into the inner diameter of each pipe, flush with its top edge. A central circular orifice of diameter \( 25.03 \pm 0.027\,\mathrm{mm} \) is milled through the HDPE free end surface. These larger PVC assemblies are then mounted over GEOM1 such that the free end of the inner GEOM1 pipe is flush with the outer HDPE free end surface, thereby forming composite release configurations for GEOM2, GEOM3, and GEOM4.

\begin{table}
\centering
\begin{tabular}{cccccc}
\hline
      & $d_s $ (mm) & $d_o$ (mm) & $h_s$ (mm) & $AR_1$ ($h_s/d_o$) & $AR_2$ ($d_o/d_s$)  \\ \hline
GEOM1 & 20.375      & 24.85      & 285        & 11.47   &     1.22   \\ \hline
GEOM2 & 20.375      & 60.33      & 285        & 4.72    &   2.96     \\ \hline
GEOM3 & 20.375      & 219.1      & 285        & 1.30    &   10.75    \\ \hline
GEOM4 & 20.375      & 406.4      & 285        & 0.70    &    19.95   \\ \hline
\end{tabular}
\caption{Geometrical dimensions and $AR_1$ and $AR_2$ of the four release geometries.}
\label{table:LG}
\end{table}

\subsection{Approach flow}

The approach flow is a developing ZPG turbulent boundary layer. Tests use \(U_\infty\simeq 2.24,\,4.48,\,6.72~\mathrm{m\,s^{-1}}\), giving \(\delta/h_s\simeq 1.81,\,1.56,\,1.45\). Boundary layer parameters and \(Re_{d_o}=U_\infty d_o/\nu\) are in Table~\ref{table:BL}. Mean and variance profiles confirm canonical behavior; the cylinder free end lies within the outer layer, and the upper bound of the logarithmic region is approximately \( 0.2-0.3h_s \).

\begin{table}
\centering
\begin{tabular}{cccccc}
\hline
\begin{tabular}[c]{@{}c@{}}$\simeq U_\infty$\\ (m~s\(^{-1}\))\end{tabular} & \begin{tabular}[c]{@{}c@{}}$ \simeq u_\tau$\\ (m~s\(^{-1}\) )\end{tabular} & \begin{tabular}[c]{@{}c@{}}$\simeq \delta_{99}$\\ (m)\end{tabular} & $\simeq \delta^+$ & $\simeq \delta/h_s$ & \begin{tabular}[c]{@{}c@{}}$ \simeq Re_{d_o}\times10^4$\\ {\footnotesize GEOM1/GEOM2/GEOM3/GEOM4}\end{tabular} \\ \hline
2.24 & 0.084 & 0.515 & 2800 & 1.81 & 0.371 / 0.901 / 3.27 / 6.07 \\ \hline
4.48 & 0.163 & 0.444 & 4820 & 1.56 & 0.742 / 1.80 / 6.54 / 12.1 \\ \hline
6.72 & 0.225 & 0.413 & 6210 & 1.45 & 1.11 / 2.70 / 9.82 / 18.2 \\ \hline
\end{tabular}
\caption{Boundary layer parameters at \(x_s = 27.54\) m and Reynolds number of cylinders.}
\label{table:BL}
\end{table}

\subsection{Scalar release}

A premixed gas (5\% methane / 95\% nitrogen by volume) is injected through GEOM1 at fixed \(u_s=1.03 \pm 3\%~\mathrm{m\,s^{-1}}\) in jet-in-crossflow configuration \citep{smith1998, Chochua2000}. The gas is supplied from a pressure-regulated compressed-gas cylinder. An in-line flow meter and insertion thermocouple measure volumetric flow rate and temperature. The effective velocity ratio \(r\) is the square root of the momentum-flux ratio

\begin{equation}
r=\left(\frac{\rho_s u_s^2}{\rho_\infty U_\infty^2}\right)^{1/2} ,
\label{eq:r1}
\end{equation}
and takes values \(r\simeq 0.46,\,0.23,\,0.16\). The analysis is restricted to low-momentum releases (\(r<r_c\)) that confine the jet to the roof-wake boundary, with \(r_c>1.5\) reported previously \citep{Hanna1982}.

\subsection{Scalar concentration measurements}

Methane concentration is measured with an {\it Aeris Technologies MIRA Ultra LDS} at 1 Hz via a 3.175 mm sampling probe on a two-axis traverse. Planes at \(x\simeq 0.64,\,1.66,\,4.19\) m (\(P1,P2,P3\)) are scanned with \(\sim\) 100 points per plane. Measurement grids vary between experimental runs to accommodate differences in plume spread, with spacing ranging from 40 to 70 mm. Each measurement point is sampled for 30\,s, with the duration constrained by the volume of the methane--nitrogen gas cylinder and the prescribed volumetric flow rate. A complete two-dimensional planar scan typically requires 2--3\,h of continuous operation. Convergence testing is performed using 10-min samples at selected locations (Appendix~\ref{appA}).

\begin{table}
\centering
\begin{tabular}{cccc}
\hline
GEOM1 & GEOM2 & GEOM3 & GEOM4 \\ \hline
$+$ & $\textendash$ & $\times$ & $\textemdash$ \\ \hline
\end{tabular}
\begin{tabular}{ccccc}
\hline
\begin{tabular}[c]{@{}c@{}}$\simeq U_\infty$\\ (m~s\(^{-1}\) )\end{tabular} & \begin{tabular}[c]{@{}c@{}}$ \simeq r $\\ \end{tabular} & \begin{tabular}[c]{@{}c@{}}$P1$ \\ ($x_1/h_s \simeq$ 2.24) \end{tabular} & \begin{tabular}[c]{@{}c@{}} $P2$ \\ ($x_2/h_s \simeq$ 5.82)\end{tabular} & \begin{tabular}[c]{@{}c@{}}$P3$ \\ ($x_3/h_s \simeq$ 14.70) \end{tabular}  \\ \hline
2.24 & 0.46~({\color{red}\textemdash}) & \tikzcircle{2pt} &  \redx{$\blacksquare$}  & \redx{$\blacktriangle$}   \\ \hline
4.48 & 0.23~({\color{blue}\textemdash}) &  \tikzcircle[blue, fill=blue]{2pt}  & \bluex{$\blacksquare$} & \bluex{$\blacktriangle$} \\ \hline
6.72 & 0.16~({\color{green}\textemdash}) &  \tikzcircle[green, fill=green]{2pt} & \greenx{$\blacksquare$} & \greenx{$\blacktriangle$} \\ \hline
\end{tabular}
\caption{Data Symbols.} 
\label{table:datasym}
\end{table}

\subsection{Scalar visualization} \label{sec:viz}

Smoke visualization is employed to qualitatively examine scalar dispersion near the cylinder free end and to track the initial trajectory of the injected scalar plume in the near field. A mixture of smoke and air is generated by combining the output of a centrifugal air blower with the discharge from a {\it ROSCO} fog machine in a large-volume plenum. This smoke-laden air is then routed to the release source and injected into the wind tunnel test section at \( u_s = 1.03 \pm 3\%\)~m~s\(^{-1}\). A {\it PCO Sensicam QE} records 289 images at 1 Hz (\(1376\times1040\), 12-bit). Ensemble averages provide qualitative near-source plume structure.

\section{The flow field of vertical wall-mounted cylinders}\label{cylinder flow}

\begin{figure}
\centering
\includegraphics[scale=0.2]{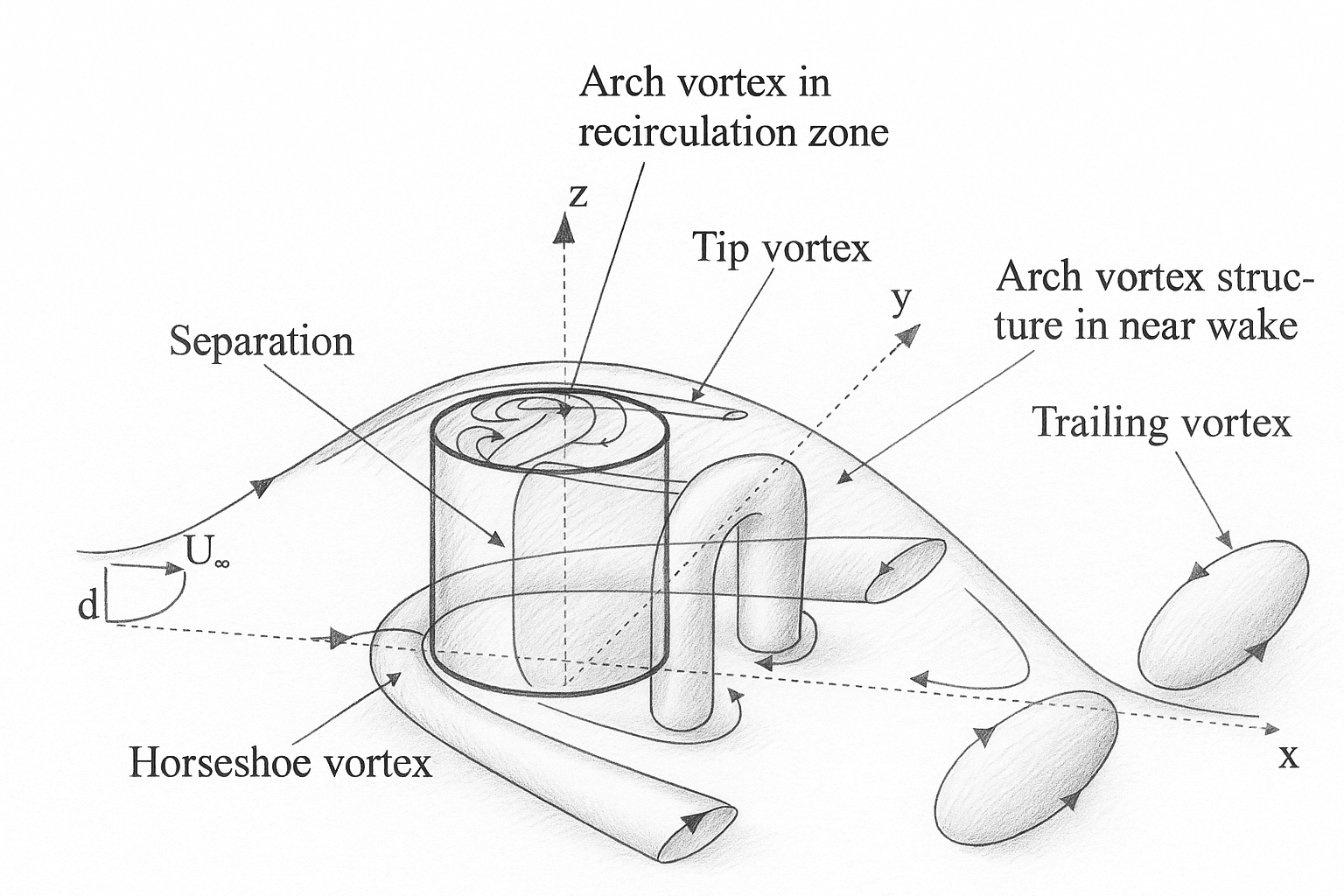}
\caption{\small Schematic of time-averaged flow over a finite-height circular cylinder (adapted from \cite{Sumner2013}).}
\label{fig:cylinder-flow}
\end{figure}

Finite-height cylinders in boundary layers produce rich three-dimensional flows with horseshoe vortices, tip vortices, arch vortices, and von K{\'a}rm{\'a}n shedding vortices \citep{Sumner2004, Pattenden2005, Palau2009, Sumner2013, Rinoshika2021, Liu2024}. The topology depends on \(Re_{d_o}\), \(AR_1\), and \(\delta/h_s\). A schematic is shown in figure~\ref{fig:cylinder-flow}.

A bifurcation with \(AR_{1,\text{crit}}\) is often reported: for \(AR_1>AR_{1,\text{crit}}\), tip and K{\'a}rm{\'a}n vortices dominate; for \(AR_1<AR_{1,\text{crit}}\), a dominant arch vortex forms in the wake, and its interaction with K{\'a}rm{\'a}n and trailing-edge vortices governs the unsteady dynamics. Literature suggests \(AR_{1,\text{crit}}\in[1,7]\), with sensitivity to \(\delta/h_s\) and configuration \citep{Sumner2013}. Dependencies include free-end reattachment, wake recirculation length, and arch vortex planform \citep{Palau2009, Rostamy2012, Rinoshika2021}.

\citet{Rinoshika2021} investigated  \( AR_1 = 1 \) and observed that the tip vortex pair is centred at \( y/d_o = \pm 0.5 \) and extends to the ground due to strong downwash near \( x/d_o = 3.5 \). A W-shaped arch vortex was identified in the wake, with the vortex head at \( y/d_o = 0 \) and shoulders at \( y/d_o = \pm 0.5 \).  The near-wake region exhibited a highly three-dimensional separated flow spanning most of the cylinder height, particularly from \( y/d_o = \pm 0.2 \). The combined effects of the W-shaped arch vortex and the horseshoe vortex on plume dispersion are discussed in \S\ref{sec:Cprimey}.

\section{Mean Scalar Fields}

\begin{figure}
\centering
\includegraphics[scale=0.4]{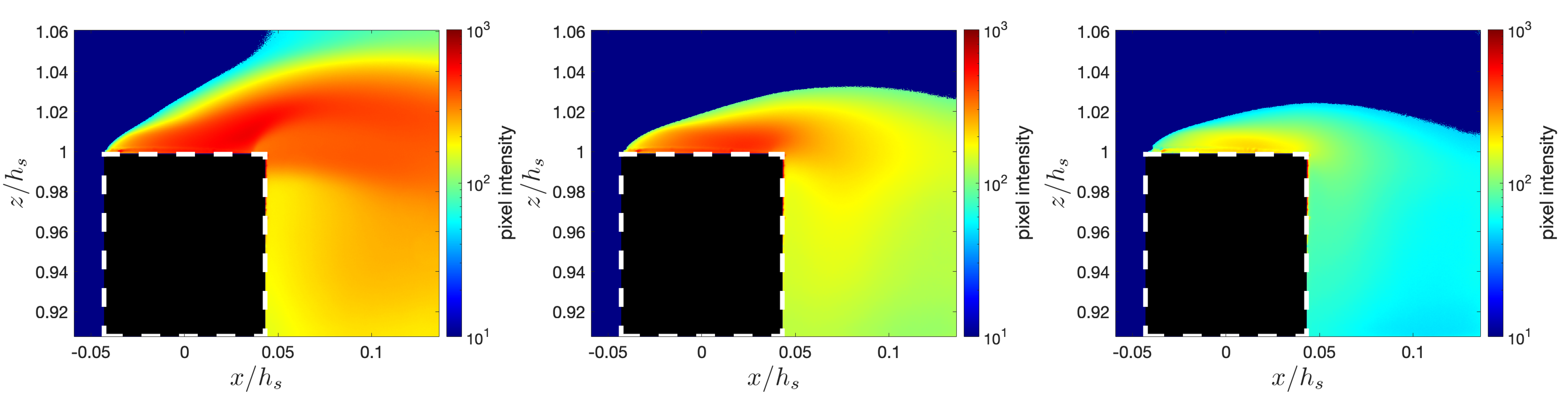}
\includegraphics[scale=0.4]{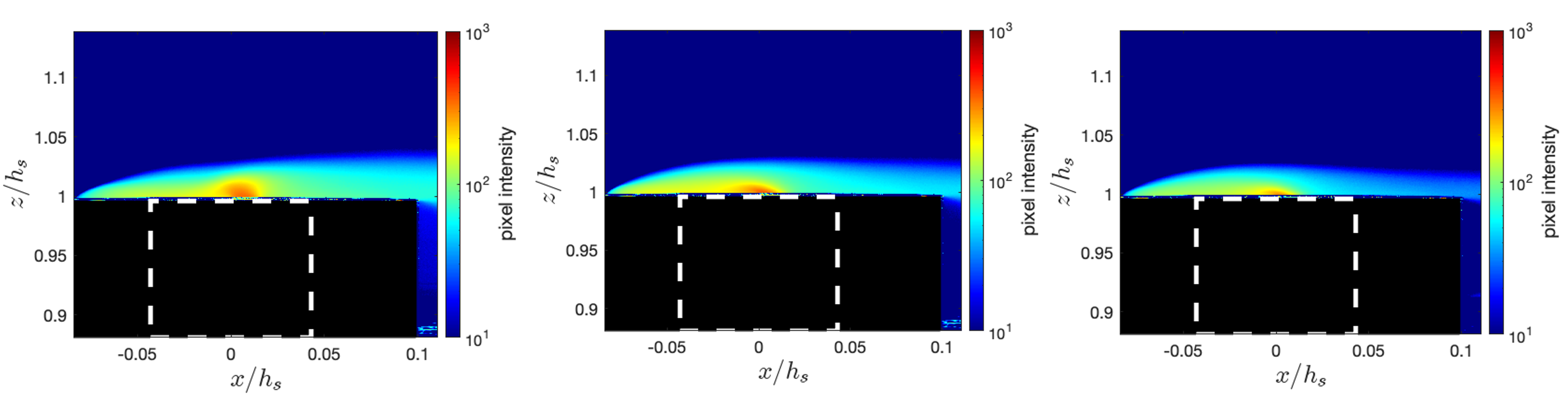}
\caption{\small Ensemble-averaged smoke images near the source: GEOM1 (top), GEOM3 (bottom). Columns: \(r\simeq 0.46, 0.23, 0.16\). Black box: outer geometry; dashed white box: source (GEOM1).}
\label{fig:FOG}
\end{figure}

\subsection{Source region}

Ensemble-averaged smoke images near the free end (\(0.9\lesssim z/h_s\lesssim 1.2\)) are shown in figure~\ref{fig:FOG} for GEOM1 and GEOM3 across \(r\simeq 0.46,\ 0.23,\ 0.16\). False-colour pixel intensity qualitatively represents scalar concentration indicating that dilution increases as \(r\) decreases. For GEOM1 the jet bends downwind and the centroid lowers with decreasing \(r\). For GEOM3, the plume extends upwind toward the leading edge due to free-end recirculation. Instantaneous images show intermittent entrainment into the wake (GEOM1) and centroid flipping between downwind- and upwind-directed (GEOM3) at 1–2 Hz. For GEOM3, the upwind-directed centroid occurs more frequently, evidenced by the higher ensemble-averaged pixel intensity near the leading edge. 

\subsection{Near and far source region}

\begin{figure}
\centering
  \includegraphics[scale=0.25]{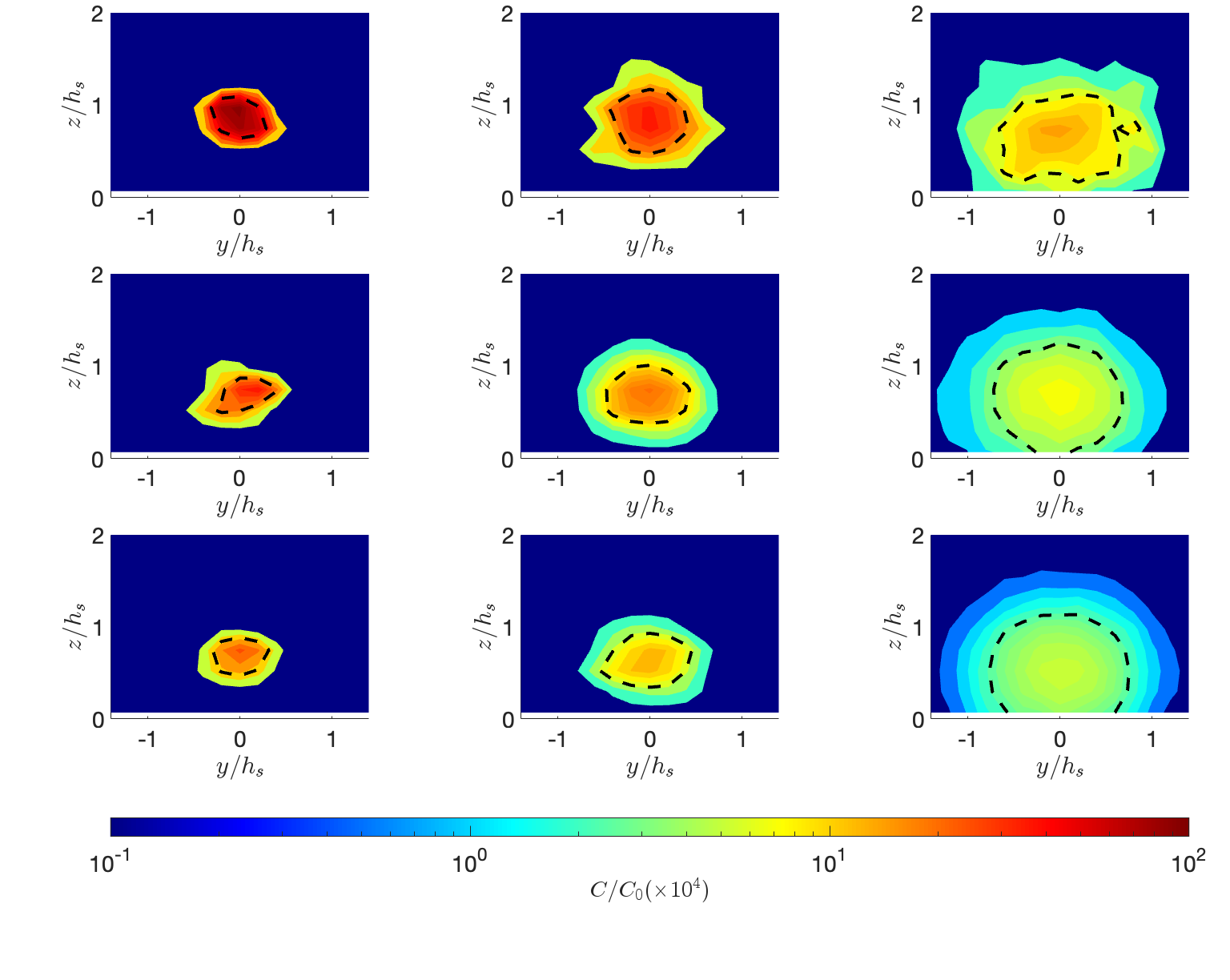}
  \vspace{-1cm}
\caption{\small{Mean concentration fields of GEOM1. Rows: \(r\simeq 0.46, 0.23, 0.16\). Columns: \(x/h_s\simeq 2.24, 5.82, 14.70\). Dashed isoline marks \(C=C_M/2\). The colour map is \( (C/C_o)\times10^4 \), where \( C_o \) is the source concentration.}}
\label{fig:CH4GEOM1}
    \vspace{0.25cm}
  \includegraphics[scale=0.175]{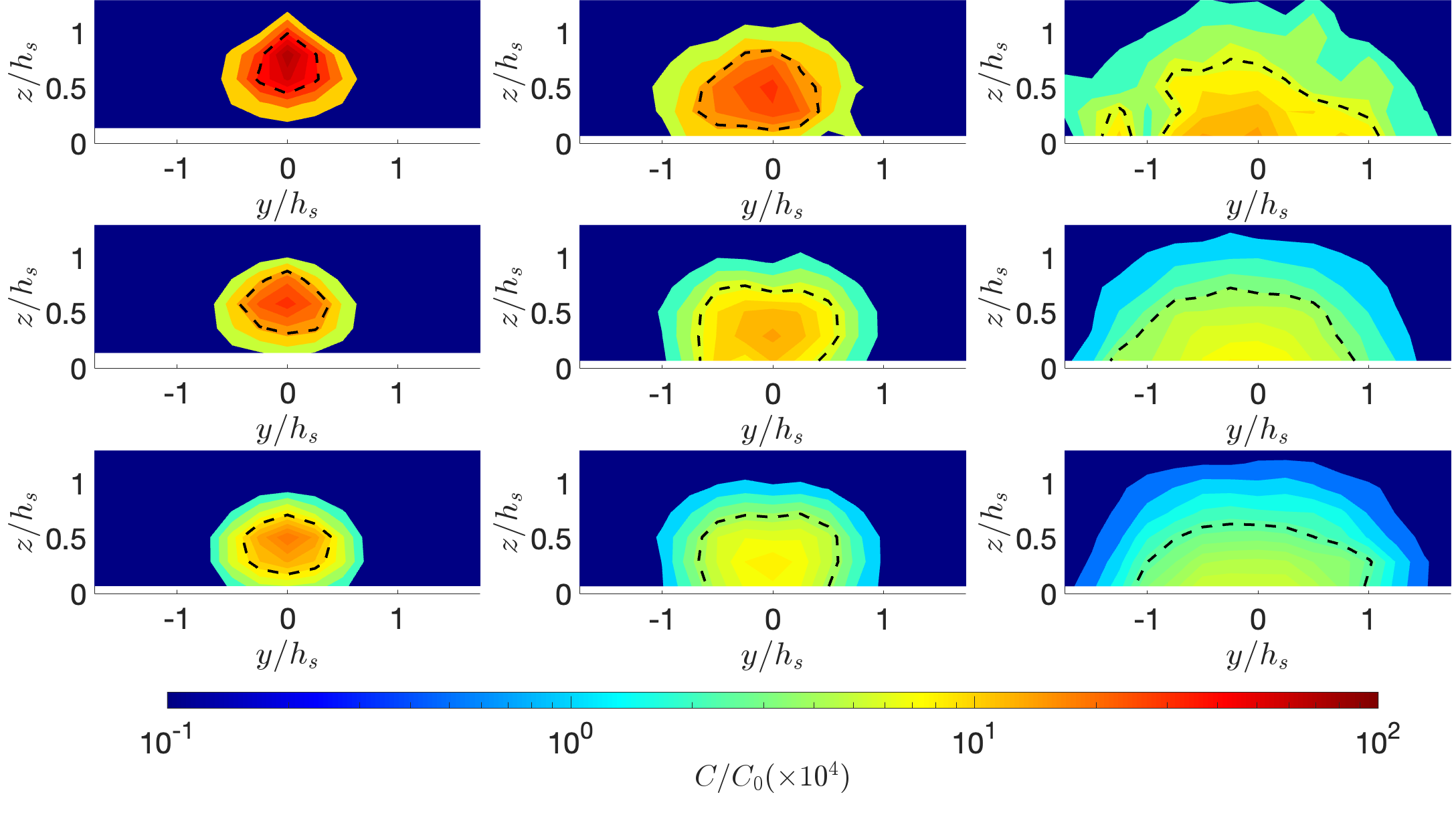}
\caption{\small{Mean concentration fields of GEOM2. The figure panel layout, colour map, and isoline are identical to those in figure~\ref{fig:CH4GEOM1}.}}
\end{figure}

\begin{figure}
\centering
\label{fig:CH4GEOM2}
\includegraphics[scale=0.175]{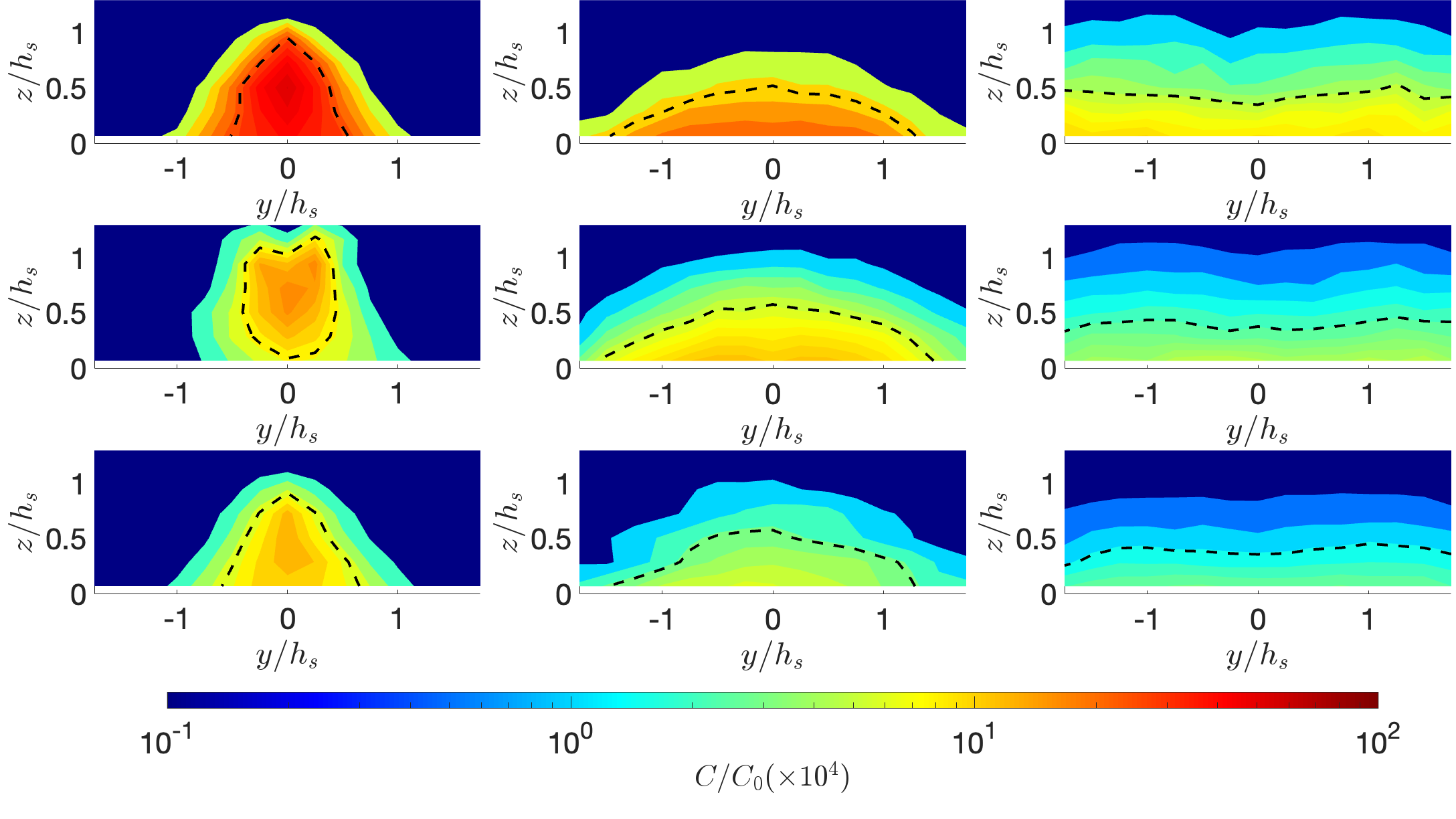}
  \includegraphics[scale=0.175]{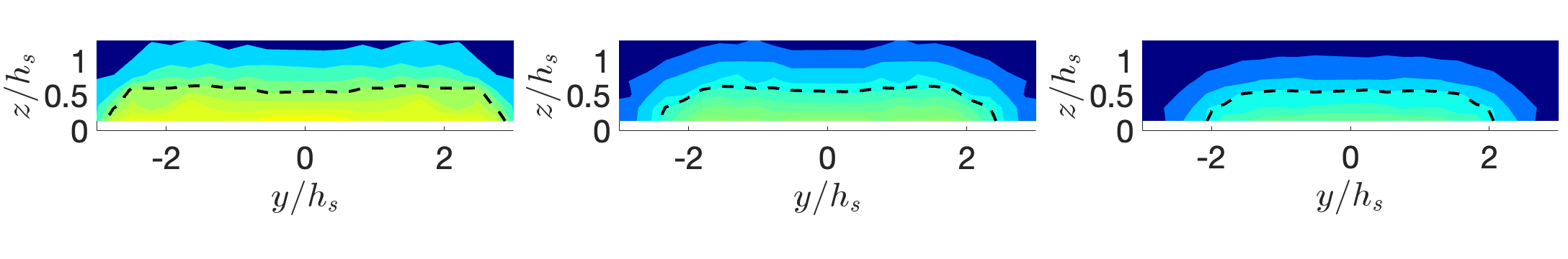}
 \vspace{-0.5cm}
\caption{\small{Mean concentration fields of GEOM3. The top \(3\times3\) panel layout, colour map, and isoline are identical to those in figure~\ref{fig:CH4GEOM1}. The bottom \(1\times3\) panel shows a wider view at \(P_3\), with \(r\) decreasing from left to right.}} 
\label{fig:CH4GEOM3}
\end{figure}

\begin{figure}
\centering
  \includegraphics[scale=0.175]{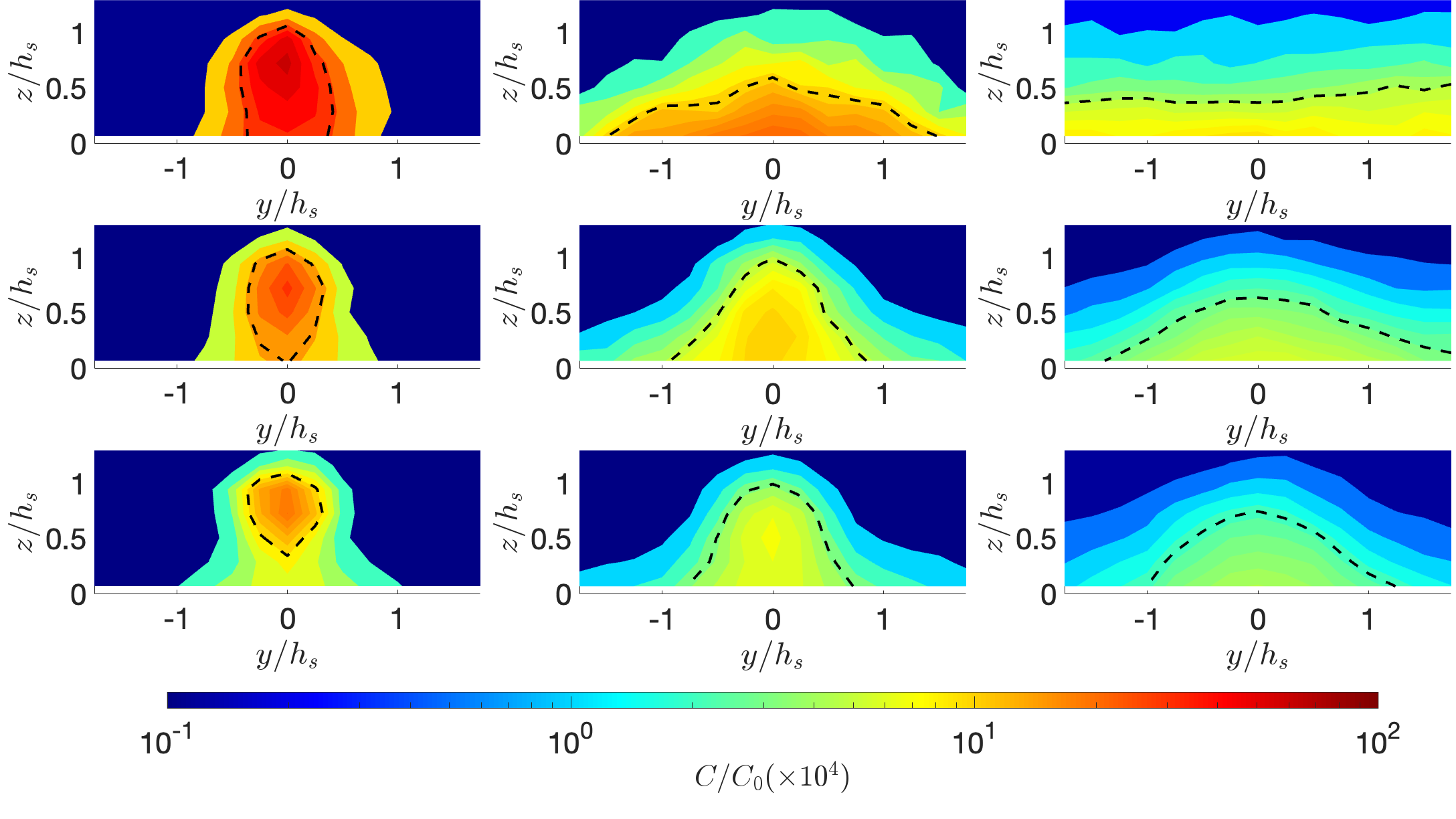}
  \vspace{-0.5cm}
\caption{\small{Mean concentration fields of GEOM4. The figure panel layout, colour map, and isoline are identical to those in figure~\ref{fig:CH4GEOM1}. The wider view of \(r1P3\) is not shown.}}
\label{fig:CH4GEOM4}
\end{figure}

Planar mean methane concentration fields are shown in figures~\ref{fig:CH4GEOM1}–\ref{fig:CH4GEOM4}. Rows: \(r\simeq 0.46, 0.23, 0.16\). Columns: \(x/h_s\simeq 2.24, 5.82, 14.70\). Dashed isoline marks \(C=C_M/2\) (with \(C_M\) the peak concentration), providing a visual estimate of the plume half-width and overall dispersion extent. The colour map is \( (C/C_o)\times10^4 \), where \( C_o \) is the source concentration. All cases show approximate lateral symmetry, stronger lateral than vertical spread, and a centroid below the source. GEOM1 behaves as an EP transitioning toward a GLP downstream; GEOM2 transitions EP--GLP--GLS across \(P1\)–\(P3\); GEOM3 and GEOM4 exhibit near-source cavity-plume features and GLS-like behavior downstream, with stronger \(r\)-dependence for GEOM4.

The mean fields highlight the dominant role of the local flow structure at the release location in shaping scalar dispersion. Aspect ratio \( AR_1 \) and velocity ratio \( r \) are key parameters. For high aspect ratios (\( AR_1 = 14.7 \), 4.72), the plume is progressively forced downward by the free-end downwash, with the EP--GLP transition point depending on both \( AR_1 \) and \( r \). For low aspect ratios (\( AR_1 = 1.3 \), 0.7), the scalar is drawn into the wake cavity, and the near-source fields resemble those of a bluff body cavity source. In the far-field, the plumes transition to GLS behavior.

\section{Measurement of the dispersion coefficients}

\subsection{Direct measurements}\label{sec:direct}

Plume growth is quantified via: (i) half-width area \(A_{hw}\); (ii) area growth rate \(\Phi=dA_{hw}/dx\); (iii) FWHM \(\delta_{\mathrm{cz}},\delta_{\mathrm{cy}}\) measured through the centroid; and (iv) HWHM \(\delta_{\mathrm{cz}},\delta_{\mathrm{cy}}\) from crosswind integration.

\begin{figure}
\centering
\includegraphics[scale=0.25]{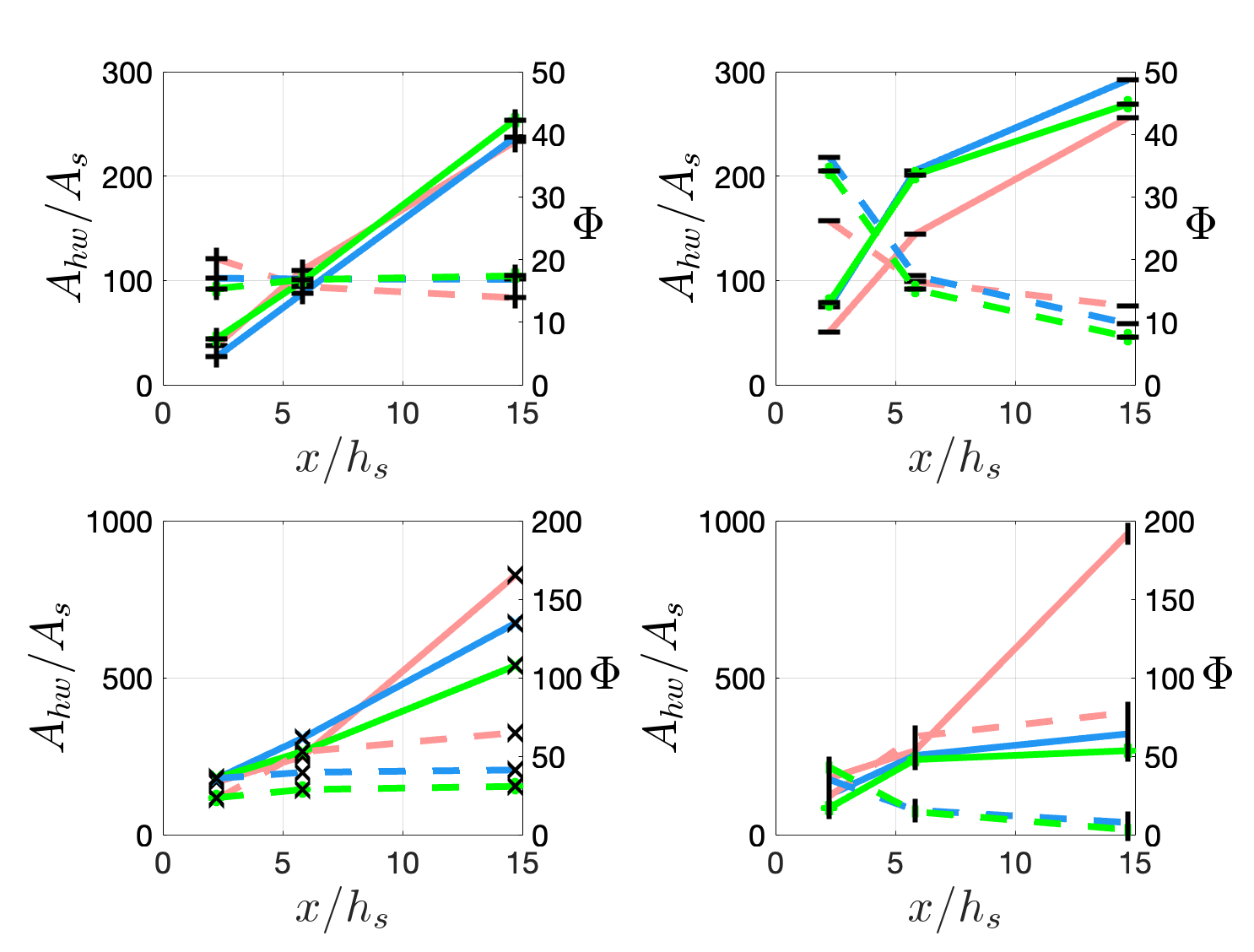}
\caption{\small Normalized plume half-width area \(A_{hw}/A_s\) (solid, left axis) and spread rate \(\Phi\) (dashed, right axis) versus \(x/h_s\). Panels (left-to-right/top-to-bottom): GEOM1–GEOM4; colours denote \(r\) per Table~\ref{table:datasym}.}
\label{fig:HWArea-Profiles}
\end{figure}

The variation of \(A_{hw}\) and \(\Phi\) with downstream position \(x/h_s\) is shown in figure~\ref{fig:HWArea-Profiles}, where \(A_{hw}\) denotes the area enclosed by the iso-contour \(C_M/2\), and \(A_s = \pi d_s^2/4\) is the source area. GEOM1 exhibits a linear increase in \(A_{hw}\), whereas GEOM2--GEOM4 display larger near-source values, reflecting enhanced lateral dispersion near the cylinder free end. GEOM2 shows an inflection at \(x/h_s \approx 5.82\), coincident with the EP--GLP transition. GEOM3 remains approximately linear except at \(r \simeq 0.46\), where \(\Phi\) increases downstream of \(x/h_s \approx 5.82\). GEOM4 exhibits the strongest \(r\)-dependence, highlighting the influence of \(AR_2\) on plume development.

\begin{figure}
\centering
\includegraphics[scale=0.47]{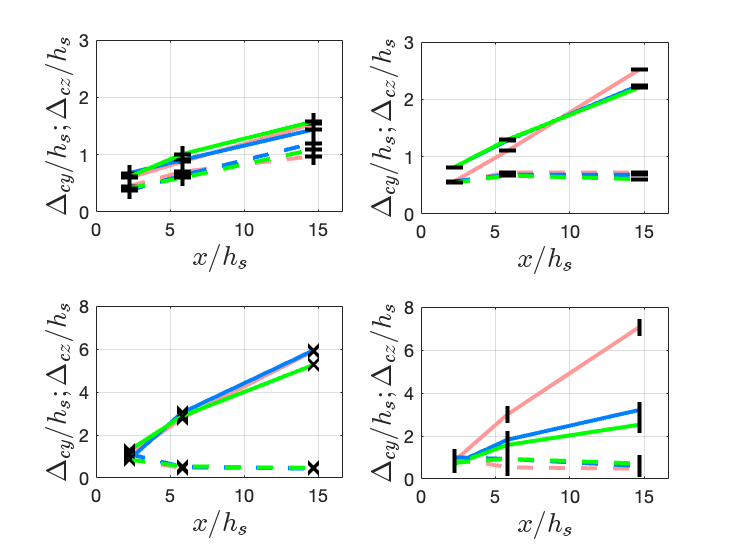}
\caption{\small FWHM \(\delta_{\mathrm{cy}}\) (solid) and \(\delta_{\mathrm{cz}}\) (dashed) versus \(x/h_s\). Panel layout and colours as in figure~\ref{fig:HWArea-Profiles}.}
\label{fig:FWHM}
\end{figure}

Full width at half maximum (FWHM) (\(\Delta_{\mathrm{cz}}\) and \(\Delta_{\mathrm{cy}}\)) are presented in figure~\ref{fig:FWHM}. The suppression of vertical plume spread associated with the EP--GLP transition is evident in the behavior of \(\Delta_{\mathrm{cz}}\). Notably, GLP and GLS plumes exhibit nearly constant vertical width, \(\Delta_{\mathrm{cz}} \approx 0.5h_s\), while the lateral extent remains significantly larger, \(\Delta_{\mathrm{cy}} \gg 2\Delta_{\mathrm{cz}}\). Together, these measures provide a quantitative indicator of plume--ground interaction.

\begin{figure}
\centering
\includegraphics[scale=0.225]{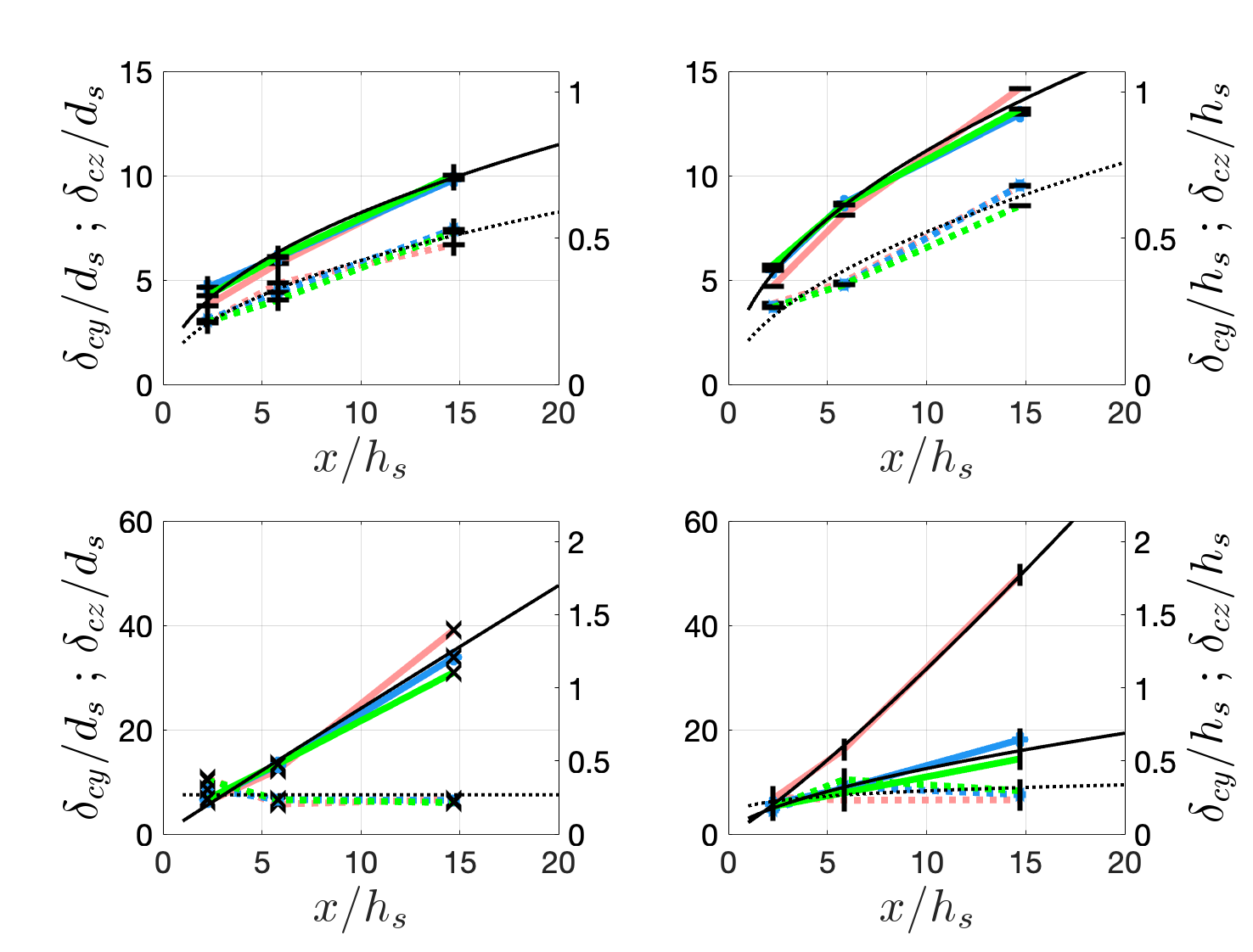}
\caption{\small HWHM \(\delta_{\mathrm{cy}}\) (solid) and \(\delta_{\mathrm{cz}}\) (dashed) versus \(x/h_s\). Black lines: power-law fits. Dual axes: left normalized by \(d_s\); right by \(h_s\).}
\label{fig:HW-Profiles2}
\end{figure}

The downstream evolution of half-width at half-maximum (HWHM) \(\delta_{\mathrm{cz}}/d_s\) (dashed) and \(\delta_{\mathrm{cy}}/d_s\) (solid) is shown in figure~\ref{fig:HW-Profiles2}. Crosswind integration suppresses the weak \(r\)-dependent variations observed in \(A_{hw}/A_s\) and the FWHM. For GEOM1, both \(\delta_{\mathrm{cy}}\) and \(\delta_{\mathrm{cz}}\) scale as \(x^{1/2}\), with \(\delta_{\mathrm{cy}}/\delta_{\mathrm{cz}} \approx 1.39 \pm 6\%\). GEOM2 also exhibits \(\sim x^{1/2}\) scaling despite reduced vertical growth during the transition, indicating that FWHM remains a more reliable metric for identifying EP--GLP transitions. GEOM3 shows \(\delta_{\mathrm{cy}} \propto x^{\sim 1}\) while \(\delta_{\mathrm{cz}}/h_s \approx 0.4\) remains nearly constant. GEOM4 transitions from \(\delta_{\mathrm{cy}} \propto x^{\sim 1}\) at \(r \simeq 0.46\) to \(\delta_{\mathrm{cy}} \propto x^{\sim 0.6}\) at lower \(r\), with \(\delta_{\mathrm{cz}}/h_s\) stabilizing in the range 0.4--0.5.

These results reflect complex interactions with vortex structures originating from the obstacle’s free end, sides, base, and shear layers. High-\(AR_1\) release geometries are dominated by tip-vortex–induced downwash, leading to a faster EP--GLP transition and suppressed vertical spread. In contrast, low-\(AR_1\) release geometries---which depend on both \(r\) and \(AR_2\)---interact strongly with wake vortices, enhancing lateral mixing while suppressing vertical dispersion. The present findings are broadly consistent with the observations of \citet{Allwine1980}, who reported reduced vertical growth (\(\sigma_z/h_s \approx 0.4\)) and lateral-to-vertical spread ratios \(\sigma_y/\sigma_z \approx 3\) for plumes released from wall-mounted structures under moderately stable stratification. Although arising from a different mechanism, a near-zero vertical growth rate, \(d\delta_{\mathrm{cz}}/dx \sim 0\), is also expected for plumes trapped beneath strong elevated-inversion caps in convective boundary layers \citep{Robson1983}. 

In addition, the results highlight the importance of \(AR_2\), which appears to depend on the ratio \(\delta_{\mathrm{cy}}/d_o\) in the near-source region. For example, at \(P1\), GEOM3 (\(\delta_{\mathrm{cy}}/d_o = 0.85 \pm 10\%\)) exhibits strong interaction with wake vortices, whereas GEOM4 shows smaller values of \(\delta_{\mathrm{cy}}/d_o = 0.48,~0.41,\) and \(0.38\) with decreasing \(r\). The markedly faster plume spread for \(r \simeq 0.46\) suggests that the plume half-width in the near-source region must exceed approximately \(0.5d_o\) to experience strong coupling with wake vortices. For narrower plumes, wake vortices exert a correspondingly weaker influence on the dispersion process.

In summary, the standard GDM scaling, \((\delta_{\mathrm{cy}}, \delta_{\mathrm{cz}} \propto x^n)\) with \(0.5 < n < 1\), does not hold for low-momentum, obstacle-influenced plumes. Accurate modeling requires the explicit inclusion of vortex-induced mixing and geometric effects, which can be captured empirically through correlations for \(\delta_{\mathrm{cy}}\) and \(\delta_{\mathrm{cz}}\) that depart from standard GDM behavior. The present results provide a foundation for developing such correlations.

\subsection{Indirect measurements}

From \eqref{eq:GDM2}, the peak concentration is

\begin{align}
C_M(x) &= \frac{Q \ln(2)}{\pi U \delta_{\mathrm{cy}} \delta_{\mathrm{cz}}} 
\left\{ \exp\left[-\ln(2) \frac{(z_M - \tilde{h}_s)^2}{\delta_{\mathrm{cz}}^2} \right] 
\left[ 1 + \exp\left(-\ln(2) \frac{4 z_M \tilde{h}_s}{\delta_{\mathrm{cz}}^2} \right) \right] \right\}.
\label{eq:MaxC-GLP}
\end{align}

\noindent
The leading prefactor in (\ref{eq:MaxC-GLP}) governs the primary decay of \(C_M(x)\). The exponential terms introduce structure associated with the EP--GLP transition that can bias inferred decay rates toward slower values if the measurement domain spans the EP--GLP regime.

\begin{figure}
\centering
\includegraphics[scale=0.6]{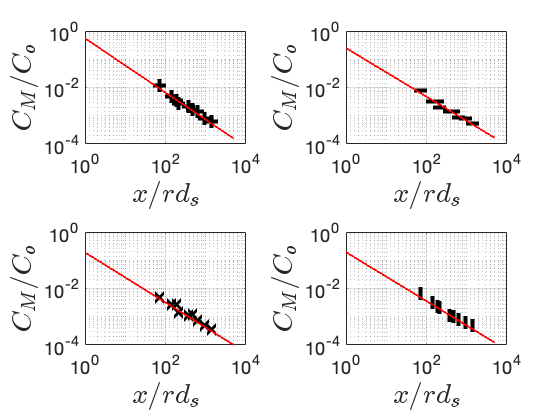}
\caption{\small{\(C_M/C_o\) versus $x/rd_s$: GEOM1 (top-left), GEOM2 (top-right), GEOM3 (bottom-left), GEOM4 (bottom-right). The solid red lines represent two parameter fits of the form $C_{M}/C_{o} = \beta (x/rd_s)^{-b}$.}}
\label{fig:Cmax}
\end{figure}

Profiles of normalized peak concentration, \(C_M/C_o\), versus \(x/(r\,d_s)\) are presented in figure~\ref{fig:Cmax}. The solid red lines denote two-parameter power-law fits of the form \(C_M/C_o = \beta (x/r d_s)^{-b}\), with fit parameters \(\beta\) and \(b\) listed in Table~\ref{tab:MaxC}.

\vspace{0.1cm}
\begin{table}
\centering
\begin{tabular}{c|cccc}
Fit parameters & GEOM1 & GEOM2 & GEOM3 & GEOM4 \\
\hline
$\beta$ & 0.55 & 0.26 & 0.22 & 0.22 \\
b & 0.94 & 0.86 & 0.89 & 0.87 \\
\end{tabular}
\caption{Best-fit relations for the normalized peak concentration \(C_M/C_o\).}
\label{tab:MaxC}
\end{table}

The sub-unity exponents (\(b<1\)) indicate that the plume undergoes EP--GLP--GLS transitions and/or that the effective dispersion coefficients \(D_y\), \(D_z\), or advection velocity \(U\) vary with \(x/r d_s\). Given \(b \approx 1\), an estimate of \(\sqrt{D_y D_z}\) can be obtained by neglecting the exponential terms in (\ref{eq:MaxC-GLP}) and substituting \(Q = C_o A_s u_s\):
\begin{equation}
\frac{C_M}{C_o} = \frac{A_s U_s}{4 \pi r d_s \sqrt{D_y D_z}} 
\left( \frac{x}{r d_s} \right)^{-1}
= \beta \left( \frac{x}{r d_s} \right)^{-1},
\label{eq:cmaxPL}
\end{equation}
where \(A_s = \pi d_s^2/4\) and \(r = u_s/U_\infty\) (neglecting small density differences between source gas and ambient air). Thus,
\begin{equation}
\beta = \frac{U_\infty d_s}{16 \sqrt{D_y D_z}},
\quad \text{or equivalently} \quad
\sqrt{D_y D_z} = \frac{U_\infty d_s}{16 \beta}.
\label{eq:beta}
\end{equation}

\noindent 
An indirect estimate of the effective diffusion coefficient product \(\sqrt{D_y D_z}\) can be obtained from~(\ref{eq:beta}). This relation is analogous to a mixing-length model in which \(D_y \sim u_y \lambda_y\) and \(D_z \sim u_z \lambda_z\), with \(u_y:u_z \propto U_\infty\) and \(\lambda_y:\lambda_z \propto d_s\).

\vspace{0.4cm}
\begin{table}
\centering
\begin{tabular}{c|cccc}
\(r\) & GEOM1 & GEOM2 & GEOM3 & GEOM4 \\
\hline
0.46 & 5.19 & 10.97 & 12.30 & 12.30 \\
0.23 & 10.14 & 21.45 & 25.35 & 25.35 \\
0.16 & 15.60 & 32.91 & 38.90 & 38.90 \\
\end{tabular}
\caption{\(\sqrt{D_y D_z}~\times10^{-3}\) (m\(^2\)~s\(^{-1}\)) computed from power-law fits.}
\label{tab:DyDz}
\end{table}

\begin{figure}
\centering
\includegraphics[scale=0.3]{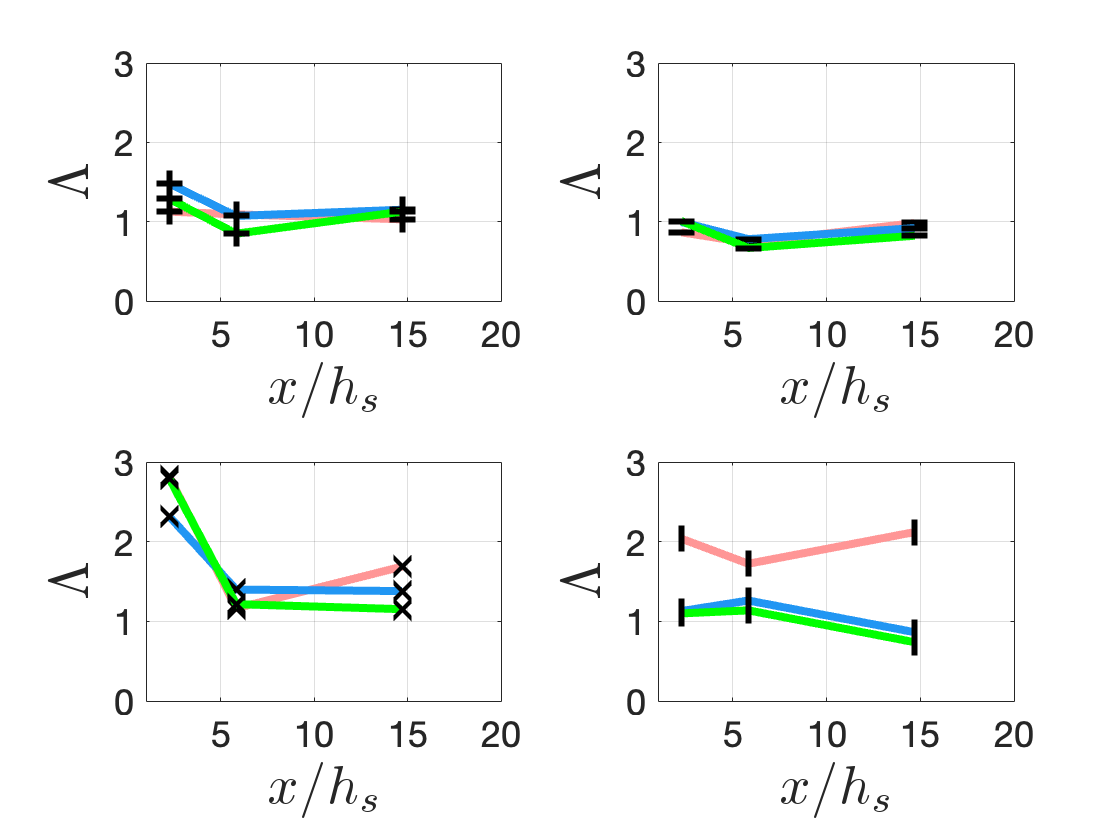}
\caption{\small{Ratio \(\Lambda=(\delta_{\mathrm{cy}}\delta_{\mathrm{cz}})_m/(\delta_{\mathrm{cy}}\delta_{\mathrm{cz}})_f\) versus \(x/h_s\) for all cases. Panels (left-to-right/top-to-bottom): GEOM1–GEOM4; colours denote \(r\) per Table~\ref{table:datasym}.}}
\label{fig:HWproduct}
\end{figure}

Assuming negligible $x$-dependence, the computed values of $\sqrt{D_y D_z}$ from Table~\ref{tab:DyDz} allow estimation of the product $\delta_{\mathrm{cy}} \delta_{\mathrm{cz}}$. These values are compared to directly measured half-width products in figure~\ref{fig:HWproduct}, using the ratio:

\[
\Lambda = \frac{(\delta_{\mathrm{cy}} \delta_{\mathrm{cz}})_m}{(\delta_{\mathrm{cy}} \delta_{\mathrm{cz}})_f},
\]

\noindent
where subscripts $m$ and $f$ denote measured and fitted values, respectively. For GEOM1 (EP) and GEOM2 (EP to GLS), the agreement is within $\pm 30\%$. For GEOM3 and GEOM4, particularly in the near-source region, the power-law fit underestimates the product by up to a factor of three.

This analysis demonstrates that the downstream decay of the peak concentration provides a valuable means for estimating the product \( \sqrt{D_y D_z} \), and consequently \( \delta_{\mathrm{cy}} \delta_{\mathrm{cz}} \). This approach is especially effective for elevated plumes, where deconvolution of the diffusion coefficients is feasible. For EPs, one can assume
\(\delta_{\mathrm{cy}}/\delta_{\mathrm{cz}}\simeq 1.4\) (see figure~\ref{fig:HW-Profiles2} discussion) to extract individual diffusion rates. For GLP/GLS, if the source height or diameter is known, the vertical half-width can be estimated as $\delta_{\mathrm{cz}} \approx 0.45 h_s$ (or roughly $6 d_s$), consistent with figure~\ref{fig:HW-Profiles2}. Moreover, power-law exponents less than 0.9 suggest that \(D_y : D_z\) decreases with $x$, indicating that the plume has encountered an upstream obstacle or undergone a transition from EP to GLP.

\section{Plume type, key GDM and WSM modeling parameter}

\begin{figure}
  \centering
  \includegraphics[scale=0.25]{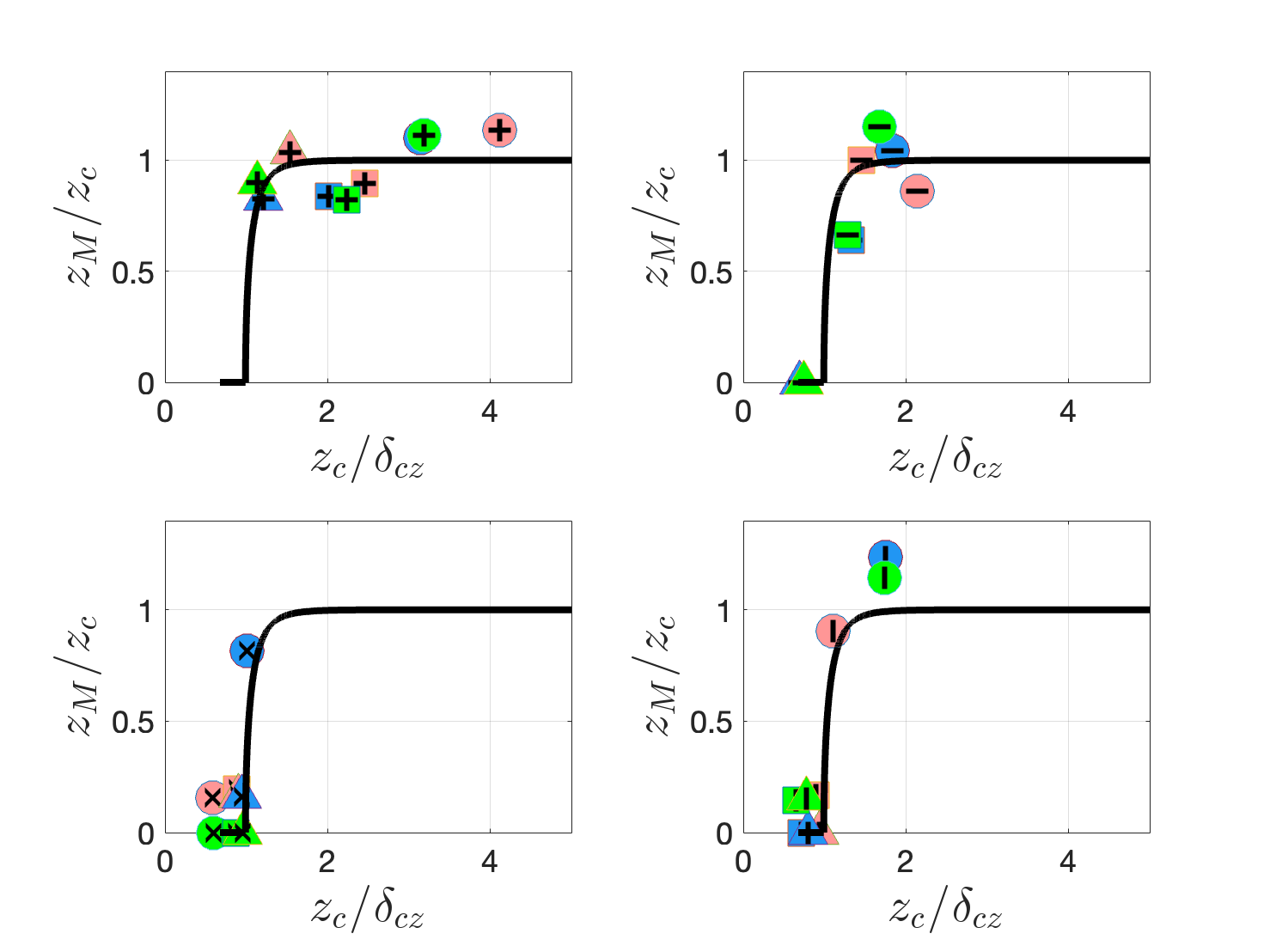}
  \caption{$z_M/z_c$ versus $z_c/\delta_{\mathrm{cz}}$. The plot provides information on the plume type in each measurement plane. The solid black line in each panel is from figure~\ref{fig:GLP}.  Panels (left-to-right/top-to-bottom): GEOM1–GEOM4; colours denote \(r\) and symbols denote \(P\) per Table~\ref{table:datasym}.}
  \label{fig:zmvszc}
\end{figure}

The plume type in each measurement plane is assessed by plotting $z_M/z_c$ versus $z_c/\delta_{\mathrm{cz}}$, as shown in figure~\ref{fig:zmvszc}. (See \S2.1, figure~\ref{fig:GLP}, and the accompanying discussion.) The solid black line in each subplot corresponds to the derived curve from the middle panel of figure~\ref{fig:GLP}, which delineates characteristic relationships for EP, GLP, and GLS plumes. The subplot layout mirrors that of figure~\ref{fig:HWproduct}; colours denote \(r\) and symbols denote \(P\) per Table~\ref{table:datasym}.

For GEOM1, the plume exhibits EP behavior in the first two measurement planes and transitions to a GLP in the final plane. GEOM2 begins as an EP at the first plane, transitions to a GLP in the second (though weaker for $r \simeq 0.46$), and becomes a GLS by the third plane, largely independent of $r$. In GEOM3, the plume already resembles a GLP or near-GLS in the first plane and clearly evolves into a GLS in the subsequent planes. GEOM4 shows similar behavior, with a GLP in the first plane and a GLS in the second and third.
The first measurement plane shows that GLP or near-GLS behavior observed for the low $AR_1$ release geometries is classified as a cavity plume, reflecting the influence of wake-induced recirculation within the obstacle’s wake cavity.

The subtle differences observed between GEOM3 and GEOM4 in the first measurement plane are likely attributable to variations in the secondary aspect ratio \(AR_2\), which modulates the influence of wake structures on the dispersion process. Overall, the regime classifications inferred from figure~\ref{fig:zmvszc} show good agreement with the visual plume behavior documented in figures~\ref{fig:CH4GEOM1}--\ref{fig:CH4GEOM4}.

\begin{figure}
  \centering
  \includegraphics[scale=0.5]{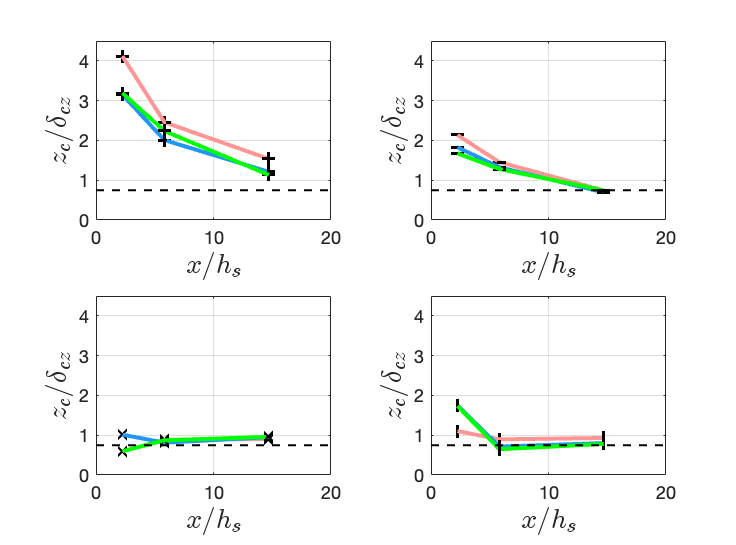}
  \caption{Normalized plume centroid $z_c/\delta_{\mathrm{cz}}$ versus $x/h_s$. The figure panel layout, symbols, and line colours are identical to those in figure~\ref{fig:HWproduct}.}
  \label{fig:centscaling}
\end{figure}

The key modeling parameter for both the GDM and WSM frameworks is the streamwise evolution of the normalized plume centroid position, \( z_c/\delta_{\mathrm{cz}} \), where \( z_c \) serves as a measurable surrogate for the effective source height, \( \tilde{h}_s \). Figure~\ref{fig:centscaling} shows \( z_c/\delta_{\mathrm{cz}} \) plotted versus \(x/h_s\). Consistent with other measures, GEOM1 and GEOM2 exhibit an \(r\)-dependence most pronounced between \(r \approx 0.46\) and \(r \approx 0.23\).  
GEOM4 shows a similar sensitivity over this \(r\) range, but with the trend reversed relative to GEOM1 and GEOM2.  

The horizontal dashed line in each subplot indicates the estimated value of \(\chi_{\mathrm{GLS}}\). As discussed in \S\ref{sec:WSM}, the analytical value of \(\chi_{\mathrm{GLS}}\) depends on \(s\) and varies slightly between the GDM and WSM formulations. The rate at which the plume approaches this asymptotic GLS structure is primarily governed by the primary aspect ratio, \( AR_1 \), with a weak dependence on \( r \) and the secondary aspect ratio \(AR_2\). Knowledge of \( z_c/\delta_{\mathrm{cz}} = f(x, AR_1, AR_2, r) \) provides essential, though not sufficient, information for accurately modeling low-momentum scalar releases from bluff bodies within a turbulent boundary layer.

\section{Characteristics of the crosswind \(y\)-integrated profiles}

The GDM crosswind \(y\)-integrated concentration profile, defined as
\( C^\prime(x,z) = \int_{-\infty}^{\infty} C(x,y,z) \, dy \), and normalized by the peak value \( C^\prime_M \), is given by:

\begin{equation}
\frac{C^\prime(x,z)}{C^\prime_M} =
\frac{ \exp \left[ -\ln(2) \frac{(z - \tilde{h}_s)^2 - (z_M - \tilde{h}_s)^2}{\delta^2_{cz}} \right] \left[ 1 + \exp \left(- \ln(2) \frac{4z\tilde{h}_s}{\delta^2_{cz}} \right) \right] }
{ 1 + \exp \left( - \ln(2) \frac{4z_M\tilde{h}_s}{\delta^2_{cz}} \right) },
\label{eq:GDM-CWY1}
\end{equation}

\vspace{0.5cm}

\noindent where all variables are defined previously. For an EP, where \( z_M = \tilde{h}_s \) and \( \tilde{h}_s/\delta_{\mathrm{cz}} > 2 \), (\ref{eq:GDM-CWY1}) simplifies to the standard Gaussian form:

\begin{equation}
\frac{C^\prime(x,z)}{C^\prime_M} = \exp \left[ -\ln(2) \frac{(z - \tilde{h}_s)^2}{\delta^2_{cz}} \right].
\label{eqn:GDM-CWY2}
\end{equation}

\noindent For a GLS, where \( \tilde{h}_s/\delta_{\mathrm{cz}} = 0 \), the profile reduces to:

\begin{equation}
\frac{C^\prime(x,z)}{C^\prime_M} = \exp \left[ -\ln(2) \left( \frac{z}{\delta_{\mathrm{cz}}} \right)^2 \right].
\label{eq:GDM-CWY3}
\end{equation}

\noindent The WSM profile, given in (\ref{eq:Cprime2}), differs from (\ref{eq:GDM-CWY3}) only in the value of the shape parameter \( s \); non-Gaussian behavior is observed when \( s \neq 2 \).

\begin{figure}
\centering
\includegraphics[scale=0.125]{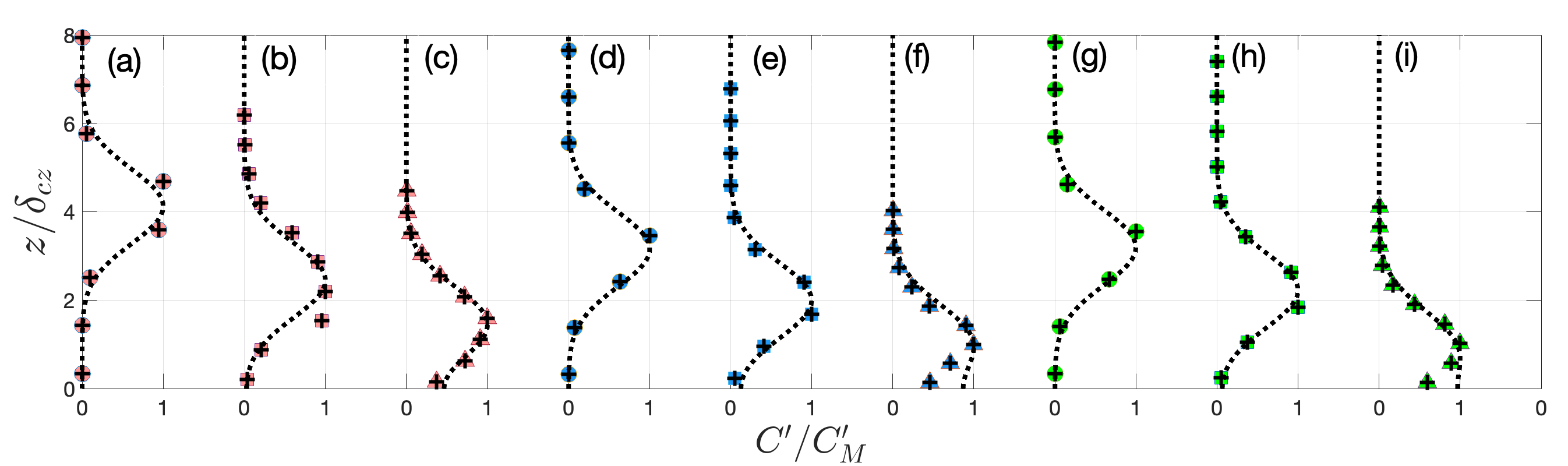}
\includegraphics[scale=0.125]{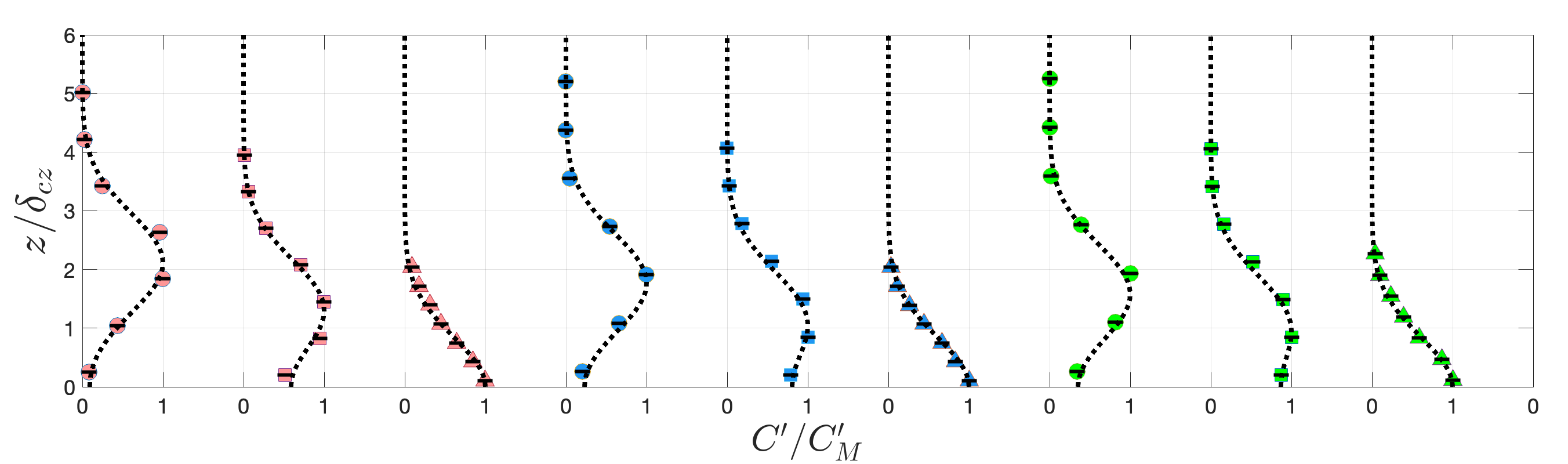}
\includegraphics[scale=0.125]{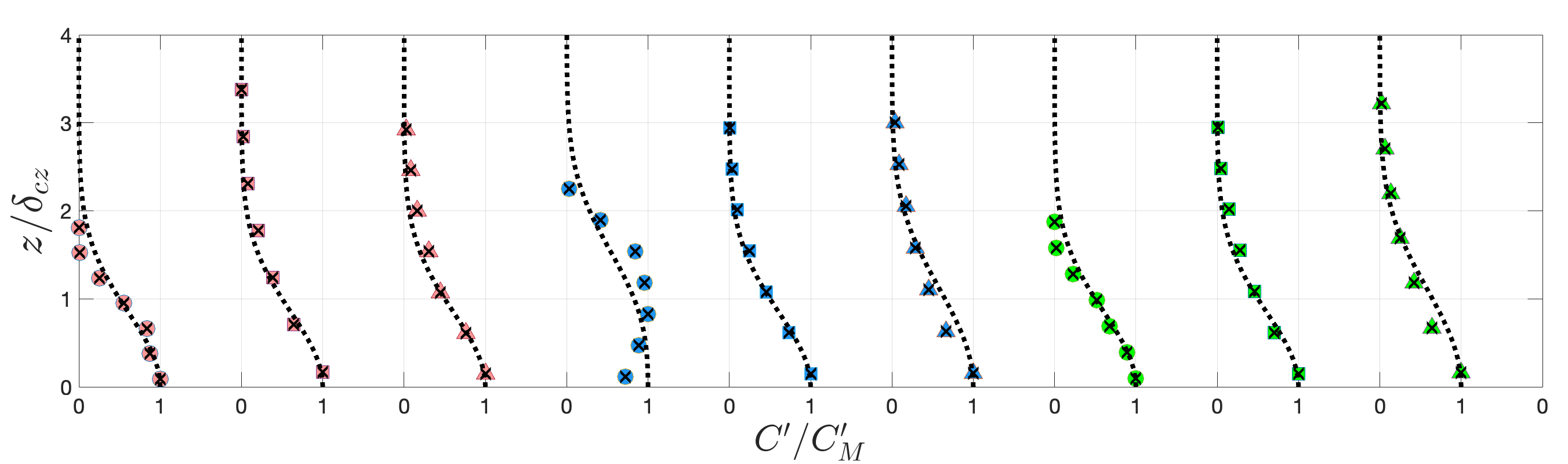}
\includegraphics[scale=0.125]{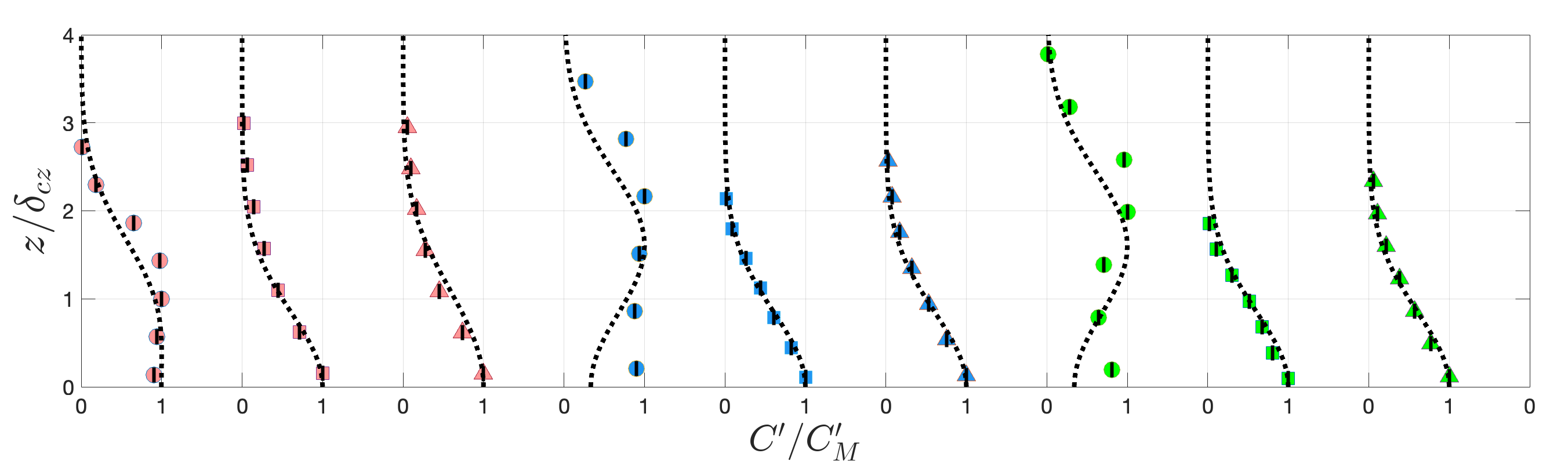}
\caption{\small{ \(C'/C'_M\) profiles for GEOM1--GEOM4 (top-to-bottom). Each panel includes nine profiles labeled (a--i), corresponding to combinations of velocity ratio and measurement plane: \( r1P1, r1P2, r1P3, \allowbreak \ldots, r3P3 \), for decreasing $r$ and increasing $P$. The symbols and colours are defined in Table~\ref{table:datasym}. The dashed lines denote the GDM profiles computed from (\ref{eq:GDM-CWY1}).}}
\label{fig:GDM-Profiles}
\end{figure}

\subsection{Assessment of GDM performance in EP--GLP--GLS regimes}

Normalized crosswind \(y\)-integrated profiles \(C'/C'_M\) for GEOM1–GEOM4 (top to bottom) are shown in figure~\ref{fig:GDM-Profiles}. Each panel includes nine profiles labeled (a--i), corresponding to combinations of velocity ratio and measurement plane: \( r1P1, r1P2, r1P3, \allowbreak \ldots, r3P3 \). The velocity ratios and measurement locations are defined in Table~\ref{table:datasym}, with \( (r1, r2, r3) \simeq (0.46, 0.23, 0.16) \) and \( (P1, P2, P3) \simeq (x = 0.64, 1.66, 4.19~\mathrm{m}) \). The dashed lines correspond to GDM predictions obtained from (\ref{eq:GDM-CWY1}--\ref{eq:GDM-CWY3}), using the measured values of \( z_c/\delta_{\mathrm{cz}} \) as input.

For GEOM1, the plume transitions from an EP at \( P1 \) to a GLP at \( P3 \) for all \( r \). The GDM provides good agreement with the data, except for profiles (f) and (i), where it overpredicts concentrations near the wall (\( z/\delta_{\mathrm{cz}} < 1 \)). For GEOM2, the plume descends more rapidly toward the wall than for GEOM1, with the transition occurring from EP (\( P1 \)) to GLP (\( P2 \)) to GLS (\( P3 \)) as \( x \) increases. The GDM fits are consistently good across all profiles and velocity ratios. The overprediction in GEOM1 (f, i) appears to stem from cases where \( z_M/\delta_{\mathrm{cz}} < 1 \), suggesting that the accuracy of the GDM is sensitive to plume behavior near the GLP–GLS transition regime. These findings suggest that the GDM effectively captures the EP–GLP–GLS transition for plumes influenced by aerodynamic downwash generated by trailing-edge vortices from the free end of the cylinder. However, it tends to overpredict wall concentrations near the GLP–GLS transition. \citet{Talluru2017} likewise documented limitations of GDM predictions for point sources positioned near the wall, corresponding to a GLP--GLS transition in the present study.

For GEOM3, the plume resembles a GLS already at \( P1 \). The response to \( r \) is non-monotonic: \( r2 \) deviates more from GLS behavior than \( r1\) or \( r3 \). By \( P2 \) and \( P3 \), the plume clearly exhibits GLS characteristics for all \( r \). GDM fits are generally strong, with mild overprediction at \( z/\delta_{\mathrm{cz}} > 1 \) for profiles (a) and (g), and moderate disagreement in (d). For GEOM4 at \(P1\), the profiles remain approximately uniform from the wall out to an \(r\)-dependent distance \(z/\delta_{\mathrm{cz}}\). This distance equals \(h_s/\delta_{\mathrm{cz}}\), indicating that \(C'\) is approximately uniform within the wake cavity. At \( P2 \) and \( P3 \), the plume fully transitions to GLS, independent of \( r \). While the GDM performs well at \( P2 \) and \( P3 \), it underperforms at \( P1 \). These results indicate that while the GDM is not well-suited for capturing the EP–GLP–GLS transition in cases dominated by full-cavity entrainment (i.e., low-\(AR_1\)), it provides accurate representations of the mean vertical concentration profiles once the plume has transitioned to a GLS.

\subsection{Comparison of GDM and WSM for GLS plumes}

\begin{figure}
    \centering

    \begin{subfigure}[b]{0.5\textwidth}
        \centering
        \includegraphics[width=\textwidth]{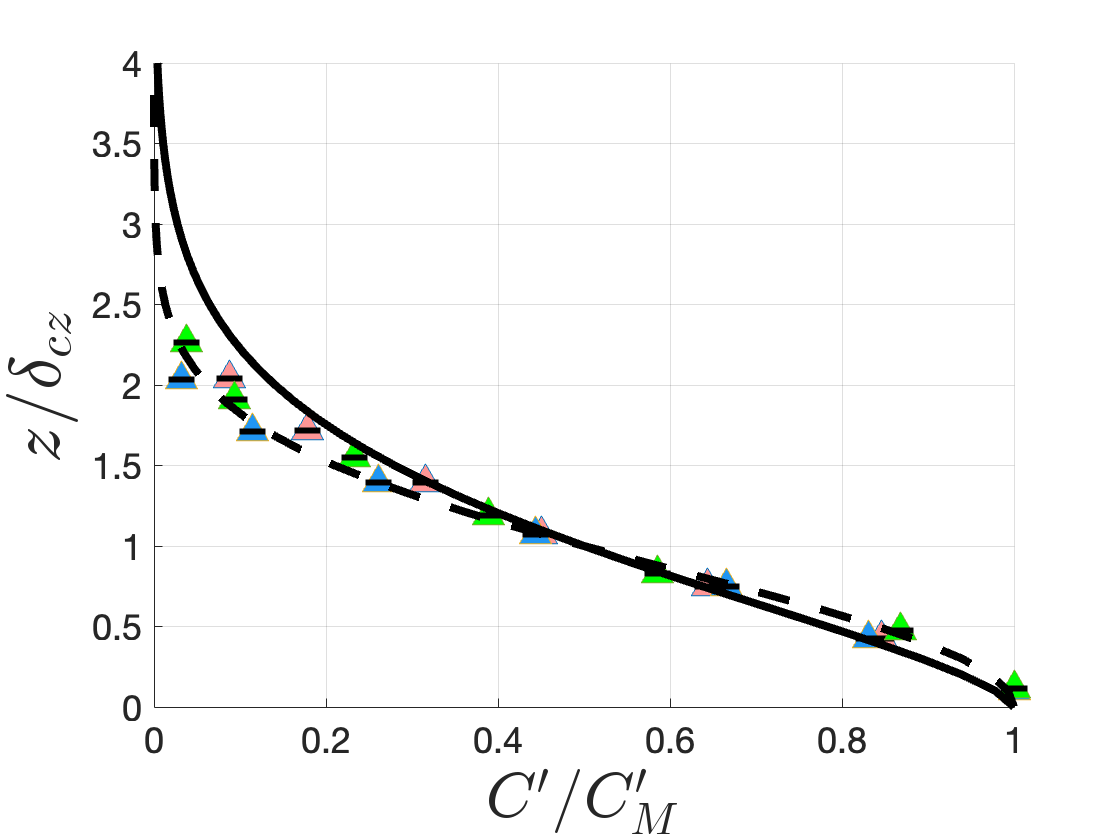}
        \caption{GEOM2: GLS profiles}
        \label{fig:top}
    \end{subfigure}

    \vspace{0.5cm} 

    \begin{subfigure}[b]{0.48\textwidth}
        \centering
        \includegraphics[width=\textwidth]{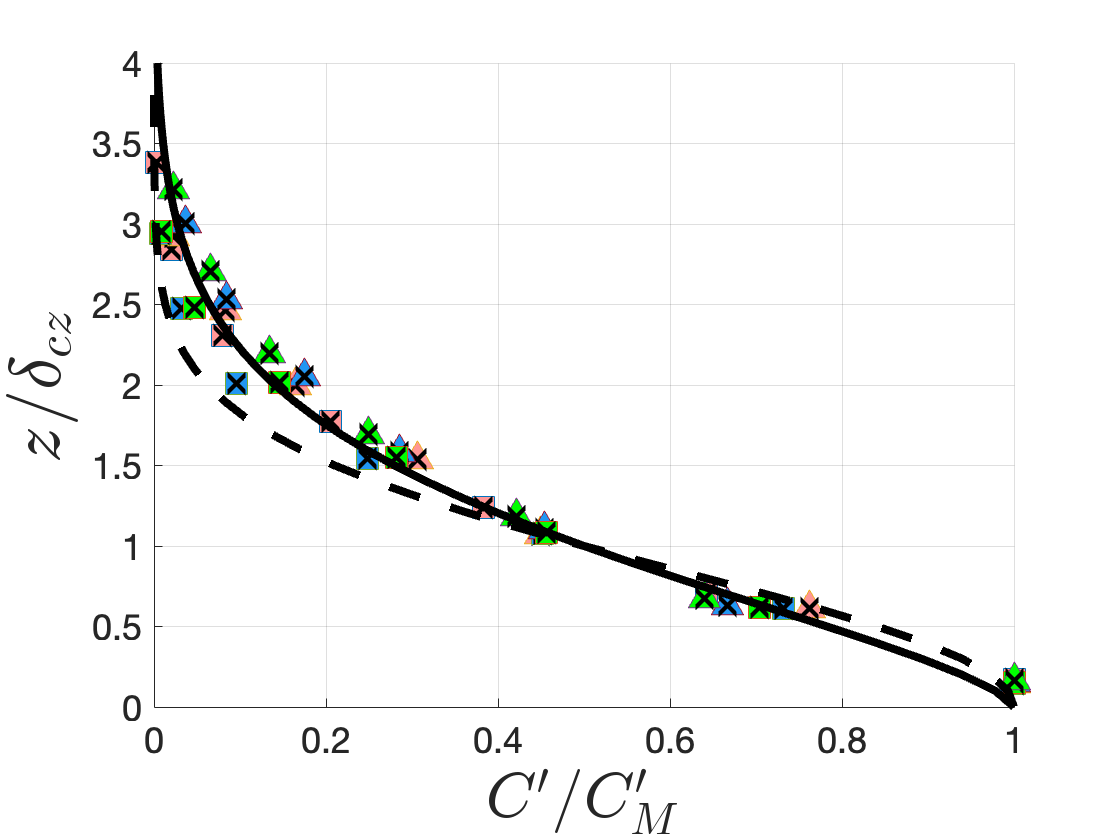}
        \caption{GEOM3: GLS profiles}
        \label{fig:bottomleft}
    \end{subfigure}
    \hfill
    \begin{subfigure}[b]{0.48\textwidth}
        \centering
        \includegraphics[width=\textwidth]{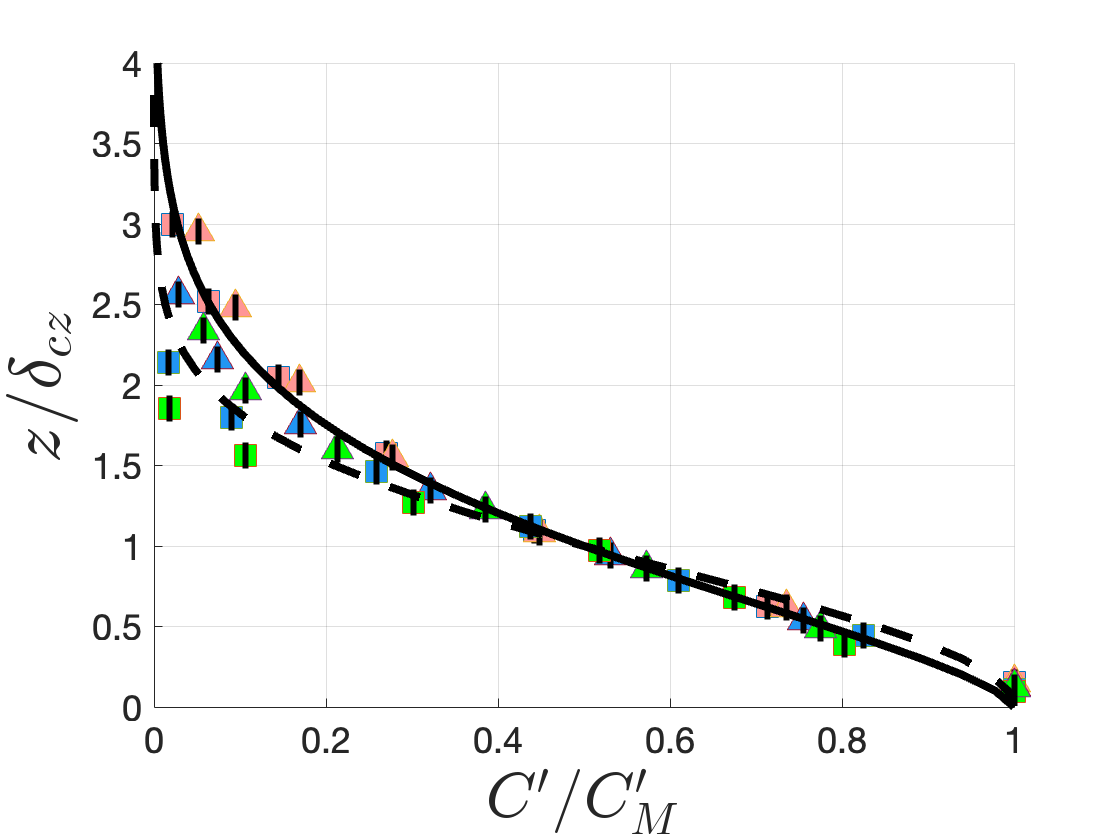}
        \caption{GEOM4: GLS profiles}
        \label{fig:bottomright}
    \end{subfigure}

    \caption{GLS profiles for (a) GEOM2, (b) GEOM3, and (c) GEOM4. The dashed line represents the GDM (\ref{eq:GDM-CWY3}), while the solid line corresponds to WSM (\ref{eq:Cprime2}) with shape exponent \( s = 1.5 \). The symbols and colours are defined in Table~\ref{table:datasym}. }
    \label{fig:triplot}
\end{figure}

The GLS concentration profiles (see figure~\ref{fig:GDM-Profiles}): GEOM2 panels (c), (f), and (i); GEOM3 panels (b), (c), (e), (f), (h), and (i); and GEOM4 panels (b), (c), (e), (f), (h), and (i), are replotted in figure~\ref{fig:triplot}. The dashed line represents the GDM prediction (\ref{eq:GDM-CWY3}), while the solid line corresponds to WSM (\ref{eq:Cprime2}) with shape exponent \( s = 1.5 \).

For GEOM2, the GLS profiles are well captured by the GDM and consistently fall below the WSM prediction. In contrast, GEOM3 profiles are more accurately represented by the WSM, with some data at \( z/\delta_{\mathrm{cz}} > 2 \) lying between the GDM and WSM predictions. GEOM4 profiles exhibit greater scatter at \( z/\delta_{\mathrm{cz}} > 2 \), particularly for \( r \simeq 0.23 \) and \( 0.16 \), although the \( r \simeq 0.46 \) profile aligns closely with the WSM.

A fit of the collective GLS datasets for each geometry yields optimal values of \( s = 2.0 \), 1.5, and 1.7 for GEOM2, GEOM3, and GEOM4, respectively. Given the difference in normalized downstream distance (\(x/d_o\)) between GEOM3 and GEOM4,  
and the proximity of the GEOM2 profiles at plane~P3 to the GLP--GLS transition, the results suggest that a finite development length beyond the GLP--GLS transition is required  
for GLS plumes to exhibit non-Gaussian behavior (\(s < 2\)).

\subsection{GDM crosswind \(z\)-integrated profiles}
\label{sec:Cprimey}
\begin{figure}
\centering
  \includegraphics[scale=0.5]{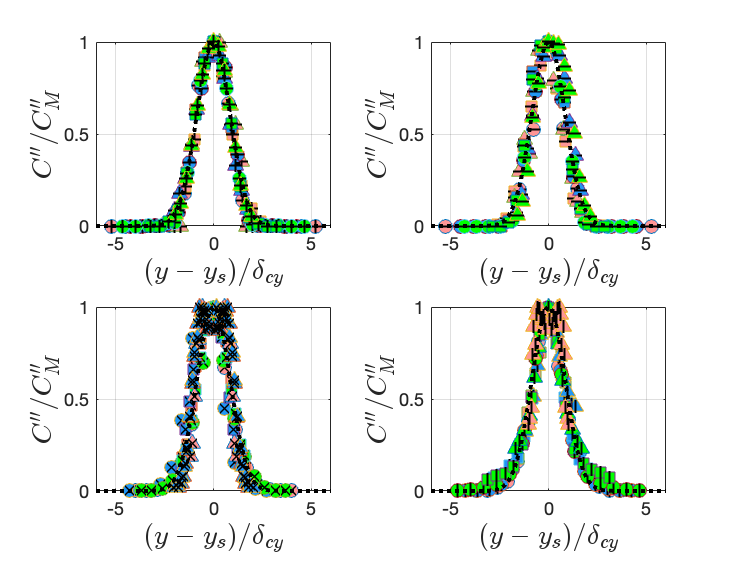}    
\caption{\small{\(C^{\prime\prime}/C^{\prime\prime}_M\) profiles. Panels (left-to-right/top-to-bottom): GEOM1–GEOM4; colours denote \(r\) and symbols denote \(P\) per Table~\ref{table:datasym}.}}
\label{fig:GLPY}
\end{figure}

The GDM crosswind \(z\)-integrated concentration profile, defined as 
\( C^{\prime\prime}(x,y) = \int_{-\infty}^{\infty} C(x,y,z) \, dz \), 
and normalized by its peak value \( C^{\prime\prime}_M \), is given by:

\begin{equation}
\frac{C^{\prime\prime}(x,y)}{C^{\prime\prime}_M} =  
     \exp \left[ -\ln(2) \frac{(y - y_s)^2}{\delta^2_{cy}} \right],
\label{eqn:GDM-EPY}
\end{equation}

\noindent where \( y_s \) is the plume centreline position and \( \delta_{\mathrm{cy}} \) is the lateral half-width. The normalized profiles are presented in figure~\ref{fig:GLPY}.

\begin{figure}
\centering
  \includegraphics[scale=0.29]{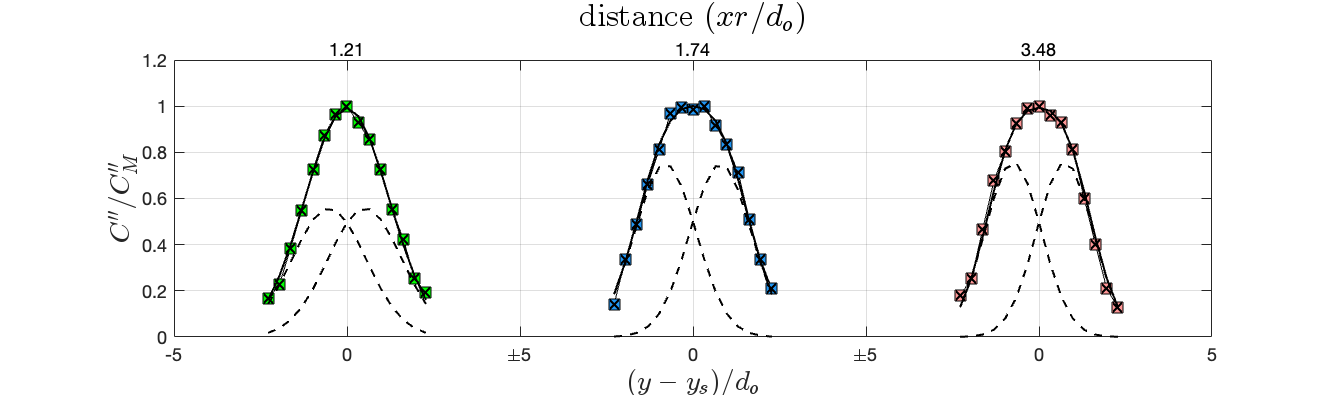} 
  \includegraphics[scale=0.29]{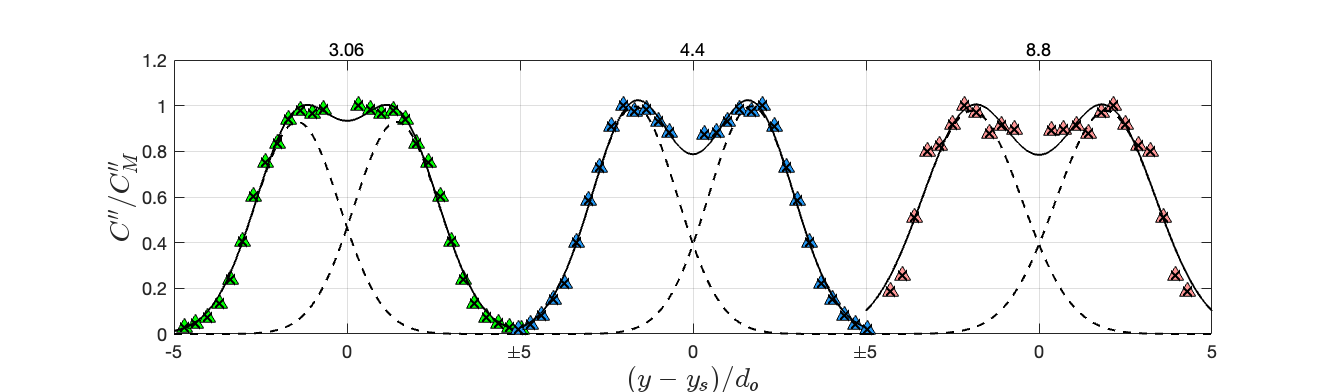} \caption{\small{\(C''/C''_M\) profiles for GEOM3 at \(P_2\) (top) and \(P_3\) (bottom), plotted against \(x/(r\,d_o)\). The solid curve equals the sum of two Gaussians (dashed).}}
\label{fig:GLPY2}
\end{figure}

For GEOM1 and GEOM2, the lateral profiles are self-similar and well-described by a Gaussian distribution across all measurement planes and for all velocity ratios \( r \). This Gaussian behavior supports the use of analytical solutions for modeling the mean concentration field \( C^\prime(x,z) \), while approximating lateral dispersion using the form in (\ref{eqn:GDM-EPY}).

In contrast, the profiles for GEOM3 and GEOM4 exhibit approximate self-similarity but deviate from Gaussian behavior in some measurement planes. Notable deviations, such as reduced peak sharpness or bimodal distributions, occur at plane $P2$ and $P3$ for GEOM3 (all \( r \)), and at plane $P3$ for GEOM4 at \( r \simeq 0.46 \). These regions coincide with those where \( \delta_{\mathrm{cy}} \propto x^{\sim1} \), indicating a link between enhanced lateral dispersion driven by wake structure interactions and non-Gaussian behavior. 

These non-Gaussian profiles for GEOM3, shown in figure~\ref{fig:GLPY2}, are well-represented by the sum of two Gaussian functions. The top row displays profiles for GEOM3 at plane $P2$, and the bottom row at plane $P3$, arranged left to right in increasing \( xr/d_o \). At the smallest \(xr/d_o\) values ($P2$, top row), the two Gaussian peaks are centred at \(y \approx \pm 0.5d_o\). As \(xr/d_o\) increases ($P3$, bottom row), the peaks spread laterally, reaching \(y \approx \pm 2d_o\) by \(xr/d_o \simeq 8.8\). This evolution suggests that within the wake cavity, scalar concentrations are initially trapped by the arch vortex near \(\pm 0.5d_o\). As the arch vortex breaks down and tilts downstream, the horseshoe vortices disperse the concentrations laterally, producing broader lateral distributions that are well represented by the sum of two symmetric Gaussians, and are thus distinctly non-Gaussian in overall form.

As described earlier, for low-\(AR_1\) release geometries, a plume half-width of at least \(\delta_{\mathrm{cy}}/d_o \gtrsim 0.5\) in the near-source region (i.e., within the wake cavity) appears to be a necessary condition for the arch vortex and the horseshoe vortex to significantly influence lateral spread.  For GEOM4, this threshold is only reached at \(r \simeq 0.46\), consistent with the observed double-peak behavior in figure~\ref{fig:GLPY} appearing only at \(r \simeq 0.46\).

\section{Data Informed Model} 

The measurement results obtained in this study suggest that a simplified dispersion model accounting for the effects of a bluff body on the EP--GLP--GLS transition is justified. The influence of the bluff body can be categorized into two distinct regimes: 
\begin{enumerate}
    \item plumes influenced by stack aerodynamic downwash (GEOM1 and GEOM2), and 
    \item plumes influenced by full cavity entrainment (GEOM3 and GEOM4).
\end{enumerate}

\noindent The key geometric parameter distinguishing these two regimes is the aspect ratio $AR_1$. Plumes dominated by stack aerodynamic downwash ($AR_1 \gtrsim 3$) undergo a gradual EP--GLP--GLS transition, characterized by vertical spreading and a decrease in the normalized plume centroid $z_c/\delta_{\mathrm{cz}}$ with downstream distance $x$, as shown in the top row of figure~\ref{fig:centscaling}. In contrast, plumes governed by full cavity entrainment ($AR_1 \lesssim 3$) transition rapidly from a uniform concentration profile within the cavity to a GLS.

\subsection{Model Construction} 

Model assumptions follow directly from those used to derive the GDM. Here, \(U(x)\), \(D_y(x)\), and \(D_z(x)\) denote the vertically averaged mean axial velocity and the effective lateral and vertical dispersion coefficients, respectively. The concentration field is assumed separable:

\[
C(x,y,z) \;=\; C_{y}(x,y)\;C_{z}(x,z)\,,
\]

\noindent where \(C_{y}\) and \(C_{z}\) satisfy the steady-state lateral and vertical advection–diffusion equations \citep{Nieuwstadt1980, Gavze2018}. In GDM form, the transverse profiles are:

\begin{align}
C_y = \exp\left[-\ln(2) \frac{(y - y_s)^2}{\delta_{\mathrm{cy}}^2} \right] 
\end{align}

\begin{align}
C_z = \beta  x^{-b}\exp\left[-\ln(2) \frac{(z - \tilde{h}_s)^2}{\delta_{\mathrm{cz}}^2} \right]
\left[ 1 + \exp\left( -\ln(2) \frac{4z \tilde{h}_s}{\delta_{\mathrm{cz}}^2} \right) \right].
\end{align}

\noindent As in most dispersion models, a primary goal is to predict the lateral and vertical dispersion widths \(\delta_{\mathrm{cy}}(x)\) and \(\delta_{\mathrm{cz}}(x)\).  Unlike standard approaches, we additionally seek to capture the EP–GLP–GLS transition, especially when it is strongly governed by fluid–structure interactions between the release geometry and the crossflow. 

\subsubsection{Fitting Functions}

Below is a concise introduction to our empirical fitting functions which, despite their data-driven form, often illuminate the underlying physics they represent.  
The detailed forms of the functions and their corresponding fit parameters are summarized in Table~\ref{tab:empiricalfits}.

\vspace{0.3cm}

\noindent {\underline{Peak concentration}}\\
The downwind evolution of the peak concentration is modeled using a power-law relationship:
\begin{equation}
\frac{C_M}{C_o} = \beta \left( \frac{x}{r d_s} \right)^{-1},
\label{mod:MaxC}
\end{equation}
where \( \beta\) is an empirical function of the aspect ratio \( AR_1 \). Specific functional forms and fitting parameters for \(\beta\) are listed in Table~\ref{tab:empiricalfits}. 

\vspace{0.3cm}
\noindent {\underline{Lateral plume half-width}}\\
A best-fit empirical power-law model for the lateral dispersion coefficient that captures the effects of \(AR_1\), \(AR_2\), and \(r\) is given by:
\begin{equation}
\frac{\delta_{\mathrm{cz}}}{d_s} = C_1 \left( \frac{x}{h_s} \right)^{\gamma_1}, \quad \text{with }
C_1 = \frac{j_1}{r^{j_2}}.
\label{mod:HWY}
\end{equation}

\noindent Fitting parameters for \(j_1, j_2\) and the specific functional form and fitting parameters for \(\gamma_1\) are listed in Table~\ref{tab:empiricalfits}

\vspace{0.3cm}
\noindent {\underline{Ratio of plume centroid to vertical half-width}}\\
The downstream evolution of the normalized plume centroid, \(z_c/\delta_{\mathrm{cz}}\), is the primary model parameter, as it governs the wall-normal centroid position and determines the streamwise locations of the EP--GLP--GLS transitions. It is modeled as:
\begin{equation}
\frac{z_c}{\delta_{\mathrm{cz}}} = (1 + \Theta)\hat{z_c}_{\mathrm{,o}}\left(\frac{x}{d_o}\right)^{-\gamma_2} + \chi_{GLS},
\label{mod:ZC}
\end{equation}
where \(\hat{z_c}_{\mathrm{,o}}= 2 \tilde{h}_s/d_s\ - \chi_{GLS}\) sets the initial condition (where \(\chi_{GLS}\) denotes the theoretical limiting value for a GLS) and \(\Theta(AR_1,AR_2,r)\) accounts for plume rise/fall in the near-source region. The exponent \(\gamma_2\) is modeled as a function of \(AR_1\). 
Specific functional forms and fitting parameters for \(\hat{z_c}_{\mathrm{o}}\) and \(\gamma_2\) are listed in Table~\ref{tab:empiricalfits}.

\begin{table}
\renewcommand{\arraystretch}{1.5} 
\setlength{\tabcolsep}{6pt}       
\centering
\caption{Empirical fitting functions and corresponding fit coefficients and fit parameters.}
\label{tab:empiricalfits}
\begin{tabular}{l|l}
\hline
\textbf{Empirical function} & \textbf{Fit coefficients and parameters} \\
\hline
\multicolumn{2}{c}{\rule{0pt}{3ex}\it{Downwind Evolution of Peak Concentration (\ref{mod:MaxC})}} \\ 
\hline
$\displaystyle \frac{C_M}{C_o} = \beta \left(\frac{x}{rd_s}\right)^{-1} $ 
&  \\ & \\
$\displaystyle \beta(AR_1) = \frac{a_1}{1 + \exp[-a_2 (AR_1 - a_3)]} + 0.22$ 
& $a_1 = 0.34, \; a_2 = 1.68, \; a_3 = 5.93$ \\
\hline
\multicolumn{2}{c}{\rule{0pt}{3ex}\it{Downwind evolution of lateral plume half-width (\ref{mod:HWY})}} \\
\hline
$\displaystyle \frac{\delta_{\mathrm{cz}}}{d_s} = C_1 (x/h_s)^{\gamma_1}, \quad C_1 = \frac{b_1}{r^{b_2}}$ 
& $b_1 = 1.88, \; b_2 = 1/3, \; \gamma_1(\alpha)$ \\
 & $\displaystyle \alpha = \frac{AR_1}{r^{AR_2/20}}$  \\
$\displaystyle \gamma_1 = LL + \frac{UL - LL}{1 + \exp[(\alpha - g_1)/g_2]}$ 
& $LL = 0.5, \; UL = 1.2, \; g_1 = 3.1, \; g_2 = 0.64$ \\
\hline
\multicolumn{2}{c}{\rule{0pt}{3ex}\it{Downwind evolution of ratio plume centroid to vertical half-width (\ref{mod:ZC})}} \\
\hline
$\displaystyle \frac{z_c}{\delta_{\mathrm{cz}}} = (1+\Theta)\hat{z_{co}} \left(\frac{x}{d_o}\right)^{-\gamma_3} + \chi_{GLS}$ 
& $\hat{z_c}_{\mathrm{co}} = 2 \tilde{h}_s/d_s -\chi_{GLS}$, $\Theta(\phi, R)$, $\gamma_2(AR_1)$  \\ & \\ 
$ \displaystyle \Theta = R\left[\mathrm{tanh}(3\phi)\right]^3$ & $\displaystyle \phi = \log(AR_1/AR_2)$ \\ & \\
$\displaystyle \gamma_2(AR_1) = j_1 + \frac{j_2 - j_1}{1 + (AR_1/j_3)^k}$ & $j_1 = 0.77$, $j_2  = 7.002$, $j_3 = 3.342$, $m = 6.902$  \\ & \\
$\displaystyle \chi_{GLS} =  B / (\ln 2)^{1/s} \approx 0.68$ for GDM, 
& $B(s) = \Gamma(2/s) / \Gamma(1/s), (1.5 \leq s \leq 2)$ \\
\hline
\end{tabular}
\end{table}

\subsection{Assessment of the model}

The essential components of a data-driven scalar-dispersion model capable of capturing the EP--GLP--GLS transition in plumes dominated by fluid–structure interactions between release geometry and crossflow have been assembled. In this section, the model is validated against experimental data for a representative set of cases. Finally, we map the predicted ground-level concentration, \(C_g\), across the \((AR_{1},AR_{2},r)\) parameter space for low-momentum releases (\(r<r_c\)).

\subsubsection{Crosswind \(y\)-integrated profiles}

\begin{figure}
\centering
\includegraphics[scale=0.3]{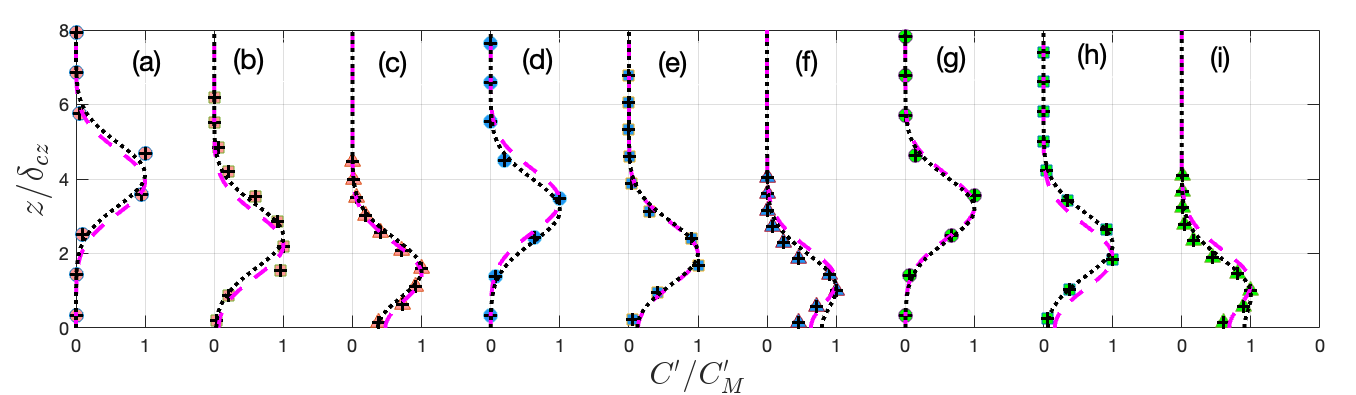}
\includegraphics[scale=0.3]{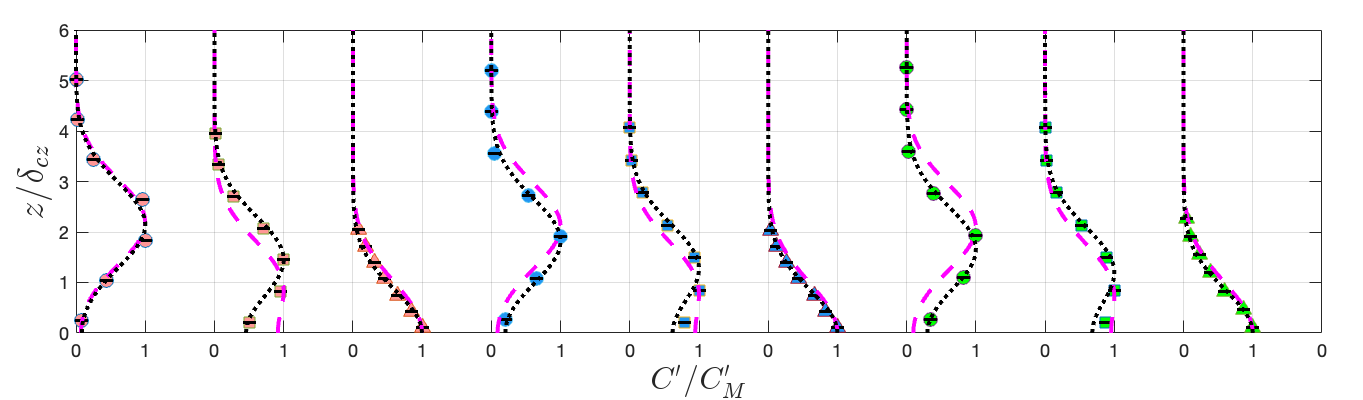}
\includegraphics[scale=0.3]{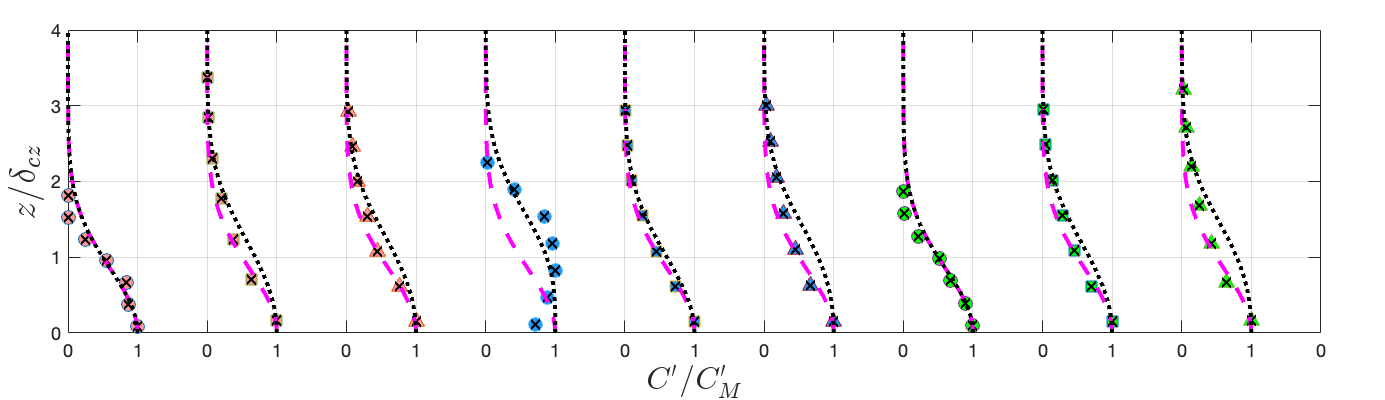}
\includegraphics[scale=0.3]{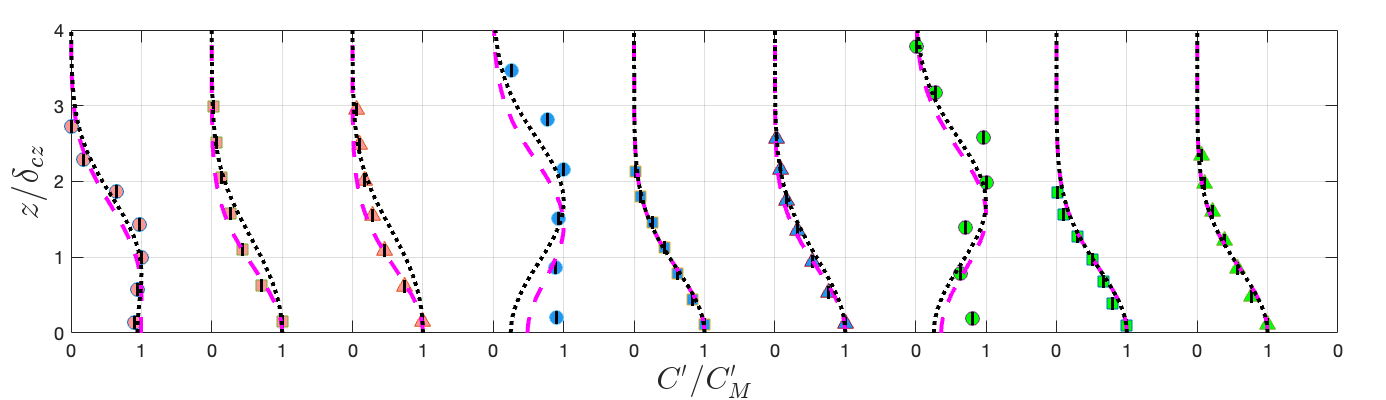}
\caption{\small{\(C'/C'_M\) profiles for GEOM1--GEOM4 (top-to-bottom). The panel arrangement matches that of figure~\ref{fig:GDM-Profiles}.  Black dashed denote GDM predictions using the measured values of \(z_c/\delta_{\mathrm{cz}}\) as inputs, while magenta dashed lines denote model results.}}
\label{fig:ModGDM-Profiles}
\end{figure}

The crosswind \(y\)-integrated concentration profiles \(C'/C'_M\) are shown in figure~\ref{fig:ModGDM-Profiles}. The panel arrangement matches that of figure~\ref{fig:GDM-Profiles}. Black dashed lines show GDM predictions from (\ref{eq:GDM-CWY1})–(\ref{eq:GDM-CWY3}), using the measured values of \(z_c/\delta_{\mathrm{cz}}\) as inputs, while magenta dashed lines denote model results.

Overall, the model reproduces the profiles for GEOM1 and GEOM2 with high fidelity, including the EP–GLP–GLS transition.  For GEOM3 the agreement remains excellent except in profile (d) ($r2P1$), where the near-uniform concentration within the wake cavity is not captured. Similarly, for GEOM4 the agreement is excellent except for profile (d) ($r2P1$) and profile (g) ($r3P1$), where the near-uniform concentration within the wake cavity is not captured. In future work, these deficiencies for low-\(AR_1\) geometries could be addressed by initializing the concentration field as uniform within the wake cavity. 

\begin{figure}
\centering
  \includegraphics[scale=0.15]{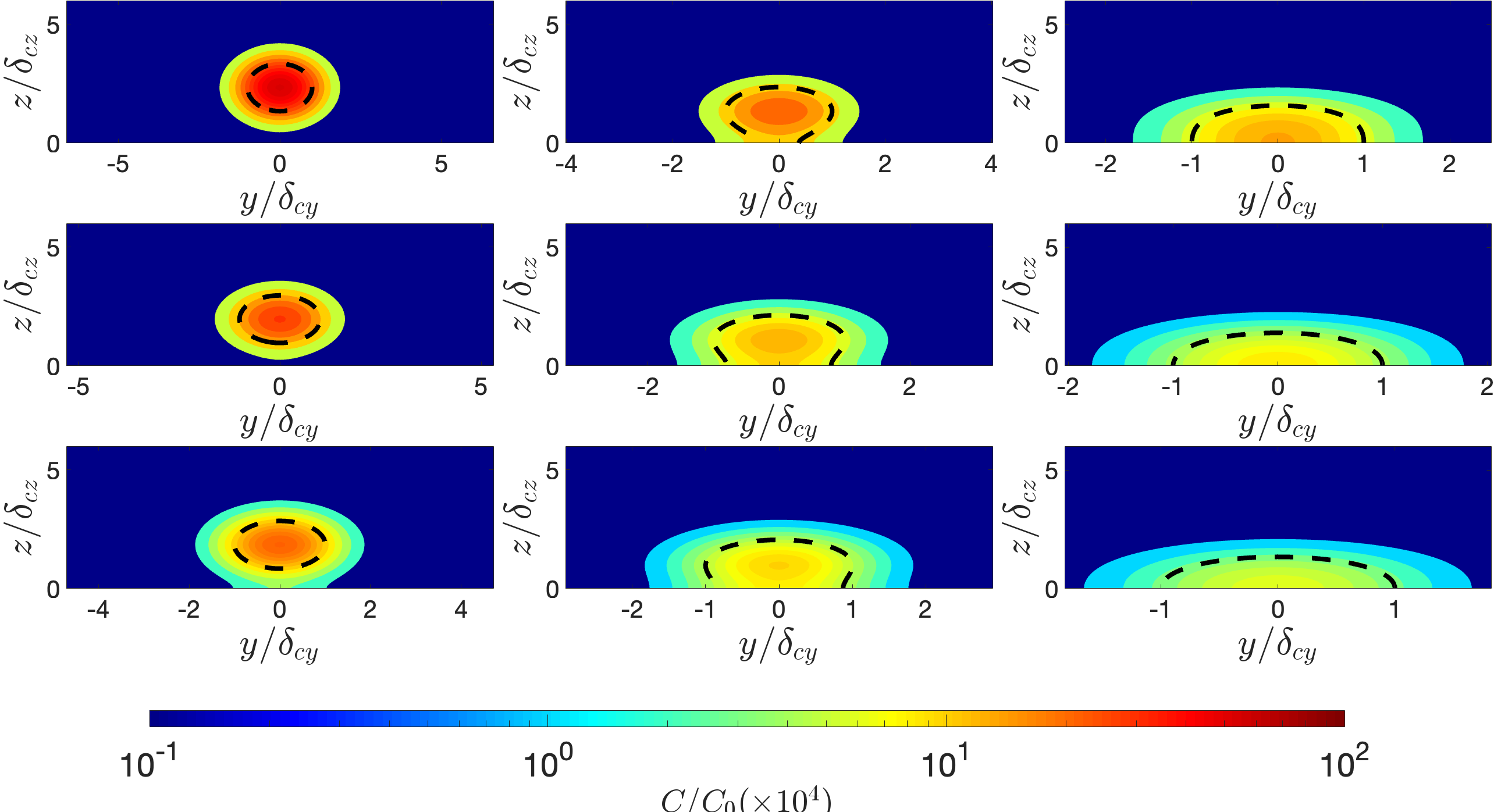}
\caption{\small{Modeled concentration fields of GEOM2. The panel arrangement is identical to figure~\ref{fig:CH4GEOM4}.}}
\label{fig:ModCH4GEOM2}
\end{figure}

\subsubsection{Two-dimensional profiles} 

Planar scalar concentration fields for GEOM2 obtained from the model are presented in figure~\ref{fig:ModCH4GEOM2}. The panel arrangement mirrors that of figure~\ref{fig:CH4GEOM4}. A key distinction between figures~\ref{fig:ModCH4GEOM2} and~\ref{fig:CH4GEOM4} lies in the coordinate scaling: in the former, the \(x\) and \(y\) axes are dynamically scaled by \(\delta_{\mathrm{cz}}\) and \(\delta_{\mathrm{cy}}\), respectively, while in the latter they are statically scaled by \(\tilde{h}_s\). Despite this difference, the model predictions exhibit excellent agreement with the measurements. First, the predicted concentration magnitudes closely match the observed values. Second, the model captures the enhanced lateral spread associated with the EP–GLP transition.

\begin{figure}
\centering
  \includegraphics[scale=0.3]{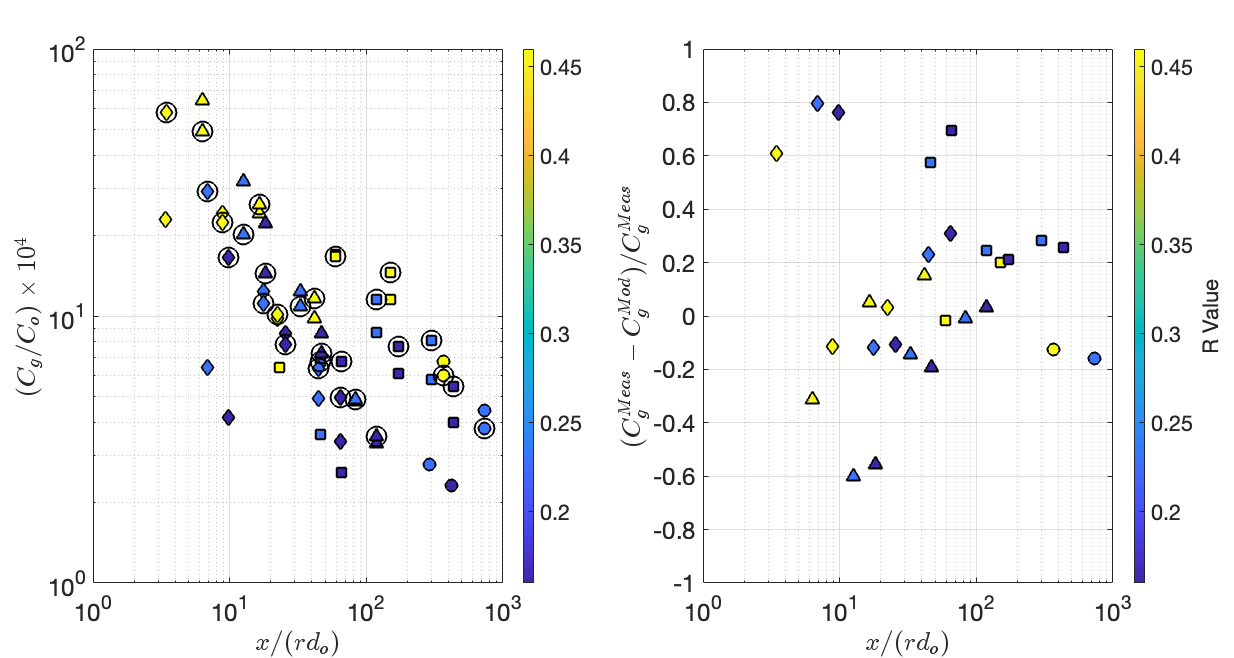}
\caption{\small{(left) Ground-level concentration, \(({C_g}/{C_o})\times10^4\) versus \(x/rd_o\) across the experimental \(AR_1\), \(AR2\), and \(r\) parameter space. The parameter \(r\) is represented by the colour map and \(AR_1\) is distinguished by the marker shape: GEOM1 (circle), GEOM2 (square), GEOM3 (triangle), GEOM4 (diamond). Encircled markers denote measurements; unencircled markers denote modeled values. (right) The normalized difference between the measured and modeled values of \(C_g/C_o\).}}
\label{fig:ModGCExp}
\end{figure}

\subsubsection{Ground-level concentration}

The model is used to predict the ground-level concentration \(C_g\) at, and downstream of, the EP--GLP transition point defined by the condition \(z_c / \delta_{\mathrm{cz}} \leq 2 \), across the \((AR_1, AR2, r)\) parameter space for low-momentum releases (\(r < 1.5\)). Measured and modeled \(C_g/C_o\) plotted against \(x/rd_o\) for the experimental \((AR_1, AR2, r)\) parameter space is shown in figure~\ref{fig:ModGCExp}. The parameter \(r\) is represented by the colour map and \(AR_1\) is distinguished by the marker shape: GEOM1 (circle), GEOM2 (square), GEOM3 (triangle), GEOM4 (diamond). Encircled markers denote measurements; unencircled markers denote modeled values.

Given that \(C_g/C_o\) requires accurate \(C_M\), \(z_c/\delta_{\mathrm{cz}}\), and \(C(z/\delta_{\mathrm{cz}})\) versus \(x/(r d_o)\), the model performs well; right-panel errors are within \(\pm 0.8\) in \(C_g/C_o\). The median absolute percentage error is \(20\%\); the root-mean-square percentage error is \(36\%\); \(76\%\) of cases fall within \(\pm 50\%\) of the measurements.

\subsection{Utility and scope of applicability}
The framework yields regime-aware, data-informed predictions of ground-level concentration and profile shape from a compact set of variables, \(C_M\) and \(z_c/\delta_{\mathrm{cz}}\), evaluated versus downwind distance \(x\). Because the diagnostics are inexpensive and tied to observables, the method is immediately useful for rapid hazard screening, experiment design, and as validation targets for LES/RANS. For modest departures in \((r,\,AR_1,\,AR_2)\) or obstacle details, the framework provides a sensible zeroth-order approximation. For larger departures, the formulation can be extended to include additional physics (e.g., strong buoyancy, unsteady crosswind, non-Boussinesq releases, or substantially different blockage/porosity). Importantly, the framework provides a clean initialization for WSM beyond the EP–GLP–GLS transition.

\section{Conclusions}

This study provides a detailed analysis of scalar dispersion from low-momentum sources released atop wall-mounted cylinders under conditions of large boundary-layer Reynolds number. By systematically varying geometric aspect ratios and the velocity ratio, we have demonstrated that plume evolution occurs through a series of well-defined transitions: from elevated plumes (EP) to ground-level plumes (GLP), and eventually to ground-level sources (GLS). These transitions are quantitatively characterized by the dimensionless parameter \( \tilde{h}_s/\delta_{\mathrm{cz}} \), with supporting metrics such as \( z_c/\delta_{\mathrm{cz}} \) and \( z_M/z_c \) providing robust plume classification.

Our results show that the EP–GLP transition is marked by suppressed vertical growth and enhanced lateral dispersion, while the GLP–GLS transition leads to near-wall scalar accumulation and asymptotic plume flattening. The influence of aspect ratio \( AR_1 \) is dominant, particularly in determining the streamwise evolution of the plume centroid, while \( AR_2 \) introduces secondary modulation by shaping the interaction between the arch vortex and the horseshoe vortex with the plume. 

Evaluation of the Gaussian Dispersion Model (GDM) and Wall Similarity Model (WSM) against the experimental data reveals that WSM more accurately captures plume behavior in GLS regimes, especially where non-Gaussian vertical profiles emerge. However, both models exhibit limitations in transition regions, highlighting the need for new or hybrid approaches.

The comprehensive dataset presented in this study provides a valuable reference for future model development and validation, including large-eddy simulations and urban dispersion models. Building on this foundation, we developed the essential components of a data-driven scalar dispersion model capable of capturing the EP–-GLP-–GLS transition in plumes dominated by fluid–structure interactions between the release geometry and crossflow. The model serves as a first-order approximation for validating numerical simulations or estimating release source locations in low-momentum releases from realistic release geometries. Beyond its practical applications, this work advances our fundamental understanding of the physical mechanisms governing scalar transport in complex, wall-bounded flows with realistic release conditions.

\appendix 
\section{}
\subsection{Measurement uncertainty in the mean methane concentration}\label{appA}

The uncertainty in the mean methane concentration measurements is an important consideration owing to the finite-time measurements acquired. The measurement uncertainty is evaluated from time-series data of methane concentration acquired over a ten-minute period on the centreline axis ($y/h_s = 0$) downstream of the cylinder. The ten-minute time-series data is divided into twenty time windows of thirty seconds each. The uncertainty in the mean methane concentration measurement is quantified by the standard deviation of the mean concentration computed over the twenty windows relative to the mean concentration computed over the ten-minute period~\citep{Shirian2023}. 

For GEOM3 and GEOM4, 10-minute time-series data were acquired at 
\[
(x/h_s, z/h_s): (2.24, 0.18), (2.24, 1.33), (5.82, 0.18), \text{ and } (5.82, 1.33).
\]
The lower and upper $z/h_s$ positions approximately correspond to the location of peak concentration and the upper plume edge, respectively. The measurement uncertainty for GEOM3 and GEOM4 is effectively identical.  
At $x/h_s = 2.24$ and $z/h_s = 0.18$ and $1.33$, the measurement uncertainty in the mean concentration for a sampling period of thirty seconds at a sampling rate of 1 Hz is $\pm 6\%$ and $\pm 21\%$, respectively.  
At $x/h_s = 5.82$ and $z/h_s = 0.18$ and $1.33$, the uncertainty is $\pm 6\%$ and $\pm 14\%$, respectively.  

The identical procedure was followed for GEOM2 with 10-minute time-series data acquired at  
\[
(x/h_s, z/h_s): (2.24, 0.5), (2.24, 0.8), (5.82, 0.5), \text{ and } (5.82, 0.8).
\]
At $x/h_s = 2.24$ and $z/h_s = 0.5$ and $0.8$, the uncertainty in the mean concentration is $\pm 5\%$ and $\pm 16\%$, respectively.  
At $x/h_s = 5.82$ and $z/h_s = 0.5$ and $0.8$, the uncertainty is $\pm 5\%$ and $\pm 13\%$, respectively.  

For GEOM1, 10-minute time-series data were acquired at 
\[
(x/h_s, z/h_s): (2.24, 0.46), (2.24, 0.98), \text{ and } (2.24, 1.33).
\]
The measurement uncertainty at these positions is $\pm 13\%$, $\pm 23\%$, and $\pm 40\%$, respectively. These measurement uncertainties are consistent with an analysis of the convergence of the running average of the time-series data, but are a factor of two larger than the 95\% confidence interval as quantified by the standard error of the mean for a thirty-second time window.  

The measurement uncertainty is largest for GEOM1, likely because the EP regime is more intermittent than GLS \citep{Nironi2015}. The sampling time required for convergence of the mean methane concentration for GEOM1 to within 10\% at $z/h_s = 0.46$, $0.98$, and $1.33$ is 60, 100, and 350 seconds, respectively. As discussed in \S2.4, these sampling times are too long owing to the finite volume of the gas cylinder. A sampling time of thirty seconds is therefore adopted as a compromise between measurement uncertainty and the ability to acquire planar measurements of mean methane concentration. Importantly, the {\it Aeris Technologies MIRA Ultra LDS} operates at \(1~\mathrm{Hz}\) and provides volume-averaged concentration over \(\approx 60~\mathrm{cm}^3\) via mid-infrared laser absorption spectroscopy. Thus, a \(30~\mathrm{s}\) sampling interval corresponds to \(\approx 1{,}800~\mathrm{cm}^3\) of sampled volume.

\subsubsection{Mean scalar flux}

A sting-mounted pitot-static tube is used to measure mean streamwise velocity at each grid point in the planar measurement grid. The pitot-static tube is sting-mounted to the two-dimensional traverse. The pitot-static tube pressure differential is measured by a {\it MKS-Baratron 10 Torr} pressure transducer. The measured mean concentration and mean streamwise velocity are used to compute the mean methane mass flux downstream of the release geometry. 

The mean streamwise velocity is measured either simultaneously with or independently of the methane concentration measurements. For simultaneous measurements, the methane measurement probe and pitot-static tube are offset by $\Delta$y, where $\Delta$y is the spanwise separation between grid points in the measurement grid. The pitot-tube measurements include corrections for the effects of viscosity, turbulence, velocity gradients, and the presence of the wall~\citep{mckeon2003}. 

The uncertainty in the mean scalar flux is a combination of the uncertainty in the methane concentration measurements and the coarseness of the measurement grid.  
The mass flux of methane, $\dot{m}$, across a $yz$-plane at an arbitrary $x$ position is

\begin{equation}
\dot{m} = \int C \, \rho_{CH_{4}} \, u \, dA,
\label{eq:flux}
\end{equation}

\noindent where \(C(y,z)\) is the measured volume fraction of methane computed from the {\it Aeris Technologies MIRA Ultra LDS}, $\rho_{CH_{4}}(P,T)$ is the density of methane dependent on the fluid pressure ($P$) and temperature ($T$), $u$ is the $x$-component of the air velocity vector (i.e., the velocity component normal to the $yz$-plane) measured with a pitot-tube, and $dA$ is an incremental area of the $yz$-plane. 

The integrated mass flux was evaluated at $x = 1.66\,\mathrm{m}$ and compared to the mass flux injected at the source, $Q$, for the four release geometries and the three freestream velocities. These twelve cases correspond to the second column in figures~\ref{fig:CH4GEOM1}--\ref{fig:CH4GEOM3}. The computed ratio $\dot{m}/Q$ for twelve nine cases has a mean of $1.04$, a standard deviation of $0.077$, a minimum of $0.90$, and a maximum of $1.18$. This demonstrates that the resolution of the measurement grid and the measurement sampling time is sufficient to evaluate the methane mass flux to within $\pm 20\%$.

\bibliographystyle{plainnat}
\bibliography{methane}

\end{document}